\documentstyle[11pt,aaspp4]{article}

\received{}
\accepted{}
\slugcomment{Submitted to the Astrophysical Journal}

\begin{document}

\def\kms{km~s~$^{-1}$ }
\def\Lya{Ly$\alpha$ }
\def\lya{Ly$\alpha$ }
\def\Lyb{Ly$\beta$ }
\def\lyb{Ly$\beta$ }
\def\Lyg{Ly$\gamma$ }
\def\Lyd{Ly$\delta$ }
\def\Lye{Ly$\epsilon$ }
\def\ang{\AA }
\def\gq{$\geq$ }
\def\lq{$\leq$ }
\def\zem{$z_{em}$ }
\def\zabs{$z_{abs}$ }

%
\title{THE COSMOLOGICAL DENSITY AND IONIZATION OF HOT GAS:\\
OXYGEN VI ABSORPTION IN QUASAR SPECTRA\altaffilmark{1}}

\author{SCOTT BURLES and DAVID TYTLER\altaffilmark{2}}
\affil{Department of Physics, and Center for Astrophysics and Space
Sciences \\
University of California, San
Diego \\
C0111, La Jolla, CA 92093-0111}

\altaffiltext{1}{Based on observations obtained with the NASA/ESA Hubble 
Space Telescope
obtained by the Space Telescope Science Institute, which
is operated by AURA, Inc., under NASA contract NAS5-26555.}
\altaffiltext{2}{scott@cass154.ucsd.edu,tytler@cass155.ucsd.edu}

%
%

\begin{abstract}
We have conducted the first survey for Oxygen VI 
$\lambda\lambda 1032,1038$ absorption lines in QSO
spectra.  We used medium resolution ($R \approx 1300$)
high signal-to-noise ($\approx 20$)
Faint Object Spectrograph spectra of 11 QSOs ($0.53\leq$\zem$\leq2.08$)
from the Hubble Space Telescope Archive.  
We use simulated spectra to determine the significance of the
line identifications, which lie exclusively in the \Lya forest.

We found 12 O~VI doublets of which 9 are expected to be real and 6 
constitute a uniform sample with both lines
exceeding a rest equivalent width of $W_r =0.21$ \AA.
The number of O~VI doublets per unit redshift at a mean
absorption redshift of $z_{ave} = 0.9$ is
$\langle N(z)\rangle = 1.0 \pm0.6$, which is similar to the density of
C~IV and Mg~II absorbers. 

In 7 of the 12 O~VI systems, O~VI, \Lyb and C~IV lines have 
similar equivalent widths, and are probably photoionized. 
In each of the remaining 5 systems, O~VI has larger equivalent widths
than those detected for \Lyb and C~IV.  These systems are labeled as
high ionization and are likely to be due to collisional
ionization. 
These would be the first QSO absorption systems known to be collisionally
ionized.

Assuming that the O~VI lines are on the linear part of
the curve of growth, we estimate the lower limit of the
cosmological mass density,
$\Omega(O~VI) \geq 1 \times 10^{-8} \, h^{-1}_{100} $.
Since O $>$ O~VI, if the
mean cosmic metallicity, Z, were below $6 \times 10^{-4}$ solar, 
then the accompanying Hydrogen and Helium
would account for all baryons in the universe. We conclude that
log $Z(z=0.9)/Z_\odot \geq -3.2$, 
and much greater if O~VI is not the
dominant ion of Oxygen.

\end{abstract}

 
%
%

\section{INTRODUCTION}

We are interested in O~VI absorption in QSO spectra
because it is the easiest way to find cosmologically distributed
hot gas with $T \approx 3 \times 10^5$~K.
Such gas may have been missed from existing surveys of QSO absorption
systems which require damped \Lya lines, Lyman continuum absorption, and
Mg~II or C~IV lines. 
Unlike other QSO absorption systems, this hot gas might be 
collisionally ionized and could comprise the bulk of all baryons.

\subsection{Source of Ionization}

We are extremely interested in whether the O~VI gas is photoionized 
or collisionally ionized.
We consider collisional ionization because
gas of the required temperatures exists, and there may be insufficient
high energy photons for photoionization, especially at low redshifts ($z \leq 1$).
We are also interested in the source of the ionizing energy -- mechanical
heating of the gas, or high energy UV photons.

Gas at temperatures needed to collisionally ionize O~VI
has been seen in galaxies, so we expect to see it in QSO absorption spectra. 
Pettini \& D'Odorico (1986) have detected
Fe~X absorption from million degree gas in the
Galactic halo and the Large Magellanic Cloud,
while Davidsen et al. (1991)
have seen O~VI in absorption from gas in the our Galaxy's halo
along the direction to 3C273. 
High latitude clouds in our Galaxy produce shadows in the
soft X-ray background (\cite{bur91,sno91})
which reveal gas at $T \sim 10^6$~K in the halo of our Galaxy, and
similar halos have been detected in other late-type galaxies (\cite{wan91}).
But the X-ray halos of early-type galaxies are too hot -- $5-23 \times 10^6$~K
-- to show strong metal lines in the HST wavelength range 
(\cite{for79,fab89}).

Photoionization is known to the dominate in most QSO absorption
systems which show strong C~IV or Mg~II (\cite{ber88}), but O~VI is
different. Photons with energies above 114~eV are needed to create O~VI, 
and these are scarce because they are absorbed by
both Hydrogen and Helium. 
QSOs and AGN are the likely sources of such photons.
QSOs and AGN apparently do emit at these energies because they show 
strong O~VI emission, and Ne~VIII ($>$207~eV) emission lines are also common
(\cite{ham95b}). But we do not know if this radiation escapes from
the QSO environment, or from the host galaxies, which might be opaque in the
He~II continuum. 

The He~II Gunn-Peterson absorption provides a measure of the spectrum of the
intergalactic background radiation field.
The IGM He~II/H~I ratio depends on the
flux of photons capable of ionizing He~II to He~III, and H~I to H~II.
Jakobsen et al. (1994) find that the optical depth at He~II
304~\AA\ is $\tau(304) \geq 1.7$, which indicates that the
spectrum of the radiation which ionizes the IGM 
is very steep.
Madua (1994) detects little flux 
above 4 Rydberg, therefore
even fewer photons exist above 114~eV. 

We conclude that the hot gas may be similar to that in the halo of 
our Galaxy and collisionally ionized 
(e.g., \cite{sem92}).
It might be heated by gas cloud collisions and supernovae explosions which
eject disk gas into the outer regions of the halo
(e.g., \cite{fan92,sha91}).
Gas heated to a million degrees cools through radiative and dielectronic
recombination on time scales of $t_c \sim 10^4 n_e^{-1} yr$,
where $n_e$ is the electron density (\cite{sha76}).
The same O~VI 2s--2p transition studied in this paper is responsible for
the large
cooling peak at $T \sim 10^5 K$ (\cite{cox69}).  To sustain 
collisionally ionized O~VI
absorption systems, the densities must
be very low ($\ll 10^{-2}$cm$^{-3}$), or the heating sources must be very
strong.

\subsection{ Density of Baryons in Hot Gas}

Hot gas with strong O~VI lines
could be the largest portion of all baryons, exceeding the
$\simeq $7\% in local stars and stellar remnants, and the
$\simeq $ 7 \% seen in cool gas as damped Ly$\alpha$ systems at
$z \simeq 3$(\cite{wol95}).  
The missing baryons might be condensed (e.g. brown dwarfs,
black holes), or diffuse hot ionized gas in clouds which are 
the subject of this paper.
Gas which is hotter than
$T \geq 10^6$~K may be seen for the first time since O~VI lines are then
stronger than those of C~IV
(Verner, Tytler \& Barthel 1994, hereafter VTB94).

The missing baryons could also be in the IGM, and too hot to show ultraviolet
lines.
COBE limits on the Compton distortion $y$ parameter limit the temperature and
density of a hot and dense IGM, but even if all baryons were in the IGM,
the maximum temperature is still a few keV (few
 $10^7$~K; \cite{wri94};\cite{bar91}),
which is still too hot to show UV absorption lines.

Many of the missing baryons could be in gas which is cooler. O~VI is most
prevalent at $T = 3 \times 10^5$~K (VTB94). By
$10^6 \leq T \leq 4 \times 10^6$~K O~VI lines will be stronger than those from
C~IV or N~V, and for highest temperatures C~IV and N~V will not be 
visible at all (VTB94).
We will only see the part of the hot gas which is enriched with metals,
This could include most hot gas if the heating is associated with
supernovae,
galaxy formation and mergers, and if all galaxies enrich the gas in their
surroundings as occur in clusters.

\subsection{Past Searches}

Although there have been no traditional 
surveys for individual O~VI lines in QSO spectra because of severe blending of
lines in the \Lya forest, the following absorption line studies have observed
O~VI and provide strong evidence that O~VI absorption is both real and common.

Hartquist \& Snijders (1982) showed that O~VI is common in 
absorption systems which are very close to QSOs, the ``associated'' C~IV
systems.  Tytler \& Barthel (unpublished) found that most associated 
systems with strong C~IV lines ($W_r (CIV) \geq 5$~\AA ) have O~VI, and 
O~VI is also common in BAL systems (\cite{ham95a}). 
In all these cases the O~VI is strong because
the absorbers are near individual QSOs.
Therefore the gas is probably photoionized,
which shows that many QSOs do emit above 114~eV.
This radiation
does escape into the QSO's galaxy, but not necessarily into the IGM.
 
Lu \& Savage (1993) formed a single composite spectrum of the absorption lines
from typical high redshift ($z_{abs}$ $\simeq 2.8$)
absorption systems which show relatively strong C~IV lines.  They found 
O~VI in systems which had C~IV, but
they could not determine the fraction of C~IV systems with O~VI.
Some had low ionization lines, while others did not. 
They did not detect N~V lines, and deduced $N$(O~VI)/$N$(N~V) $\geq 4.4$,
which implies $T \geq 2.5 \times 10^5$~K for collisional ionization.
 
Bahcall et al. (1993) note that O~VI is seen in three of the metal line 
systems in their HST quasar spectra, one of which was an associated absorber.
Bergeron et al. (1994) note that 4 of 5 metal systems
show O~VI, but they are unable to determine if the gas is collisionally
ionized or photoionized. 
These results all suggest that O~VI systems are as common 
or more common than C~IV systems (several per QSO).
We will now determine how common.

First we show that we can identify O~VI doublets in the \Lya forest.
We choose to work with HST UV spectra because the forest is 
less dense at the lower redshifts of HST spectra compared to the higher
redshifts visible from the ground. 
 We measure the frequency of O~VI systems, and we try to determine the level of
ionization, and whether the gas is photoionized or collisionally ionized.
Finally we estimate the contribution that hot gas makes to the cosmological 
mass density of baryons. 

\section{OXYGEN~VI SEARCH}

The Hubble Space Telescope Archive provides a resource to
search for O~VI absorption.  In particular, the medium resolution 
(R $\approx$ 1300)
gratings of the Faint Object Spectrograph have both the sensitivity
and the resolving power to identify weak metal lines in absorption.
Spectra are chosen by the QSO properties (magnitude and redshift), 
gratings, and exposure time.  Since the search for O~VI is done
exclusively in the Lyman-$\alpha$ forest, high signal-to-noise 
is necessary.

To ensure the reality of O~VI absorption, strict rules are
applied to line lists.  These rules incorporate
measured wavelength, equivalent width, and 1$\sigma$ errors.
Once the ground rules are determined, a computer algorithm is
responsible for identifying systems which exhibit O~VI absorption.
We use a computer to automatically find O~VI and to maintain objectivity
and repeatability. Physical significance is measured by applying the
same rules to false sets of rest wavelengths and comparing
results with the true set.  

\subsection{Processing the HST Spectra}

We consider spectra from only the FOS high resolution gratings G130H, 
G190H and G270H. The G160L does not have enough resolution, and there
were no suitable GHRS G140L spectra.  Table 1 shows the 11 QSOs chosen
for the O~VI search.
Most of the spectra are from the G190H or
G270H gratings, because few G130H spectra have adequate S/N.
There are 8 steps prior to the O~VI search,

\begin{enumerate}
\item We use the following rules to ensure that the 
$z_{em}$ of each QSO is sufficient to place O~VI well onto the
FOS gratings: 
\begin{displaymath}
z_{em} \geq \cases{0.3 &for G130H \cr 0.65 &for G190H \cr 1.25 &for G190H.}
\end{displaymath}

\item We obtained each spectrum and its error array from the HST Archive.

\item Coadd spectra which have multiple exposures.

\item Find all the absorption lines from the interstellar medium of our Galaxy.
We shift each spectrum so that these lines give a mean velocity of zero
(\cite{sch93}).  We list the offsets added to each spectrum in Table 1.

\item Fit a continuum to each spectrum manually.  
Due to the subjectivity of continuum fitting, the systematic
errors introduced 
in the measured equivalent widths are difficult to calculate.
However, we assume that other errors in our final results
greatly outweigh the errors in
the continuum fits.

\item Convert the spectrum and its error array into a 
spectrum which gives the minimum
equivalent width of unresolved lines of specified significance level.

\item Find all $5 \sigma$ lines in a spectrum. There are typically 40 -- 80.
Measure their equivalent widths $W$, $\sigma (W)$, and wavelengths.

\item Search the literature for all known absorption systems.
Identify lines which belong to these systems, and remove them
from the sample in which we will search for O~VI.

\end{enumerate}

\subsection{Search for O~VI Lines}

There are no published systematic searches for individual spectral lines in the
Ly$\alpha$ forest region because the large number of forest lines and the
high incidence of line blending lead to numerous false identification. 
We require a search which has
the desirable attributes of a medical screen for a rare disease:
(1) high sensitivity to minimize the number of missed O~VI systems
(false negatives), and (2) high specificity to minimize the number of false
positive identifications. We use a computer code to increase the sensitivity,
accuracy, completeness and speed of the survey, 
and use rules to limit the number of
false positives to about 20\% of the sample. 
We would prefer to have fewer
false positives, but this is only possible if we consider just the very
strongest lines.
Then we miss most O~VI lines, and we cannot estimate 
their overall frequency.

False identifications can be noise features, or real
lines with other identifications, especially Ly$\alpha$. We 
use three criteria to limit the number of false identification:
line significance, redshift agreement, and equivalent width ratios.  
In all detections, both O~VI lines must be detected above the
$2 \sigma$ equivalent width limit.  Furthermore, their 
measured equivalent width ratio must lie within $1 \sigma$ of 
the theoretical boundaries.

The procedures we use to find O~VI lines in certain and
possible redshift systems are described below.

\subsubsection{Search for O~VI Lines in Certain Redshift Systems}

In published, or new metal line systems (e.g. with Mg~II, C~IV
or Lyman limits) which are considered certain, 
we look for $2 \sigma$ O~VI lines. There are about 0 -- 3 such
systems per QSO, so the chance of 
 false positive identification of both O~VI lines is  
$\leq 3\%$ per QSO.  We measure lines associated with the absorption 
systems 
and calculate the mean redshift, which can vary from the
published absorption redshift because of wavelength offsets.
We identify all additional lines, including O~VI, 
which match the criteria of the next section.

\subsubsection{Search for O~VI Lines in Uncertain Redshift Systems}

We use the following procedure to
find interesting lines, identify possible new redshift systems and then reject 
unreasonable ones.

\bigskip
\begin{enumerate}
\item 
We search for all systems with at least two $5\sigma$ lines from 
Ly$\alpha$, Ly$\beta$, Ly$\gamma$ or O~VI.  Lines are accepted into a system if 
their
redshift differs from the weighted mean $z_{ave}$ by less than 0.0005 ($\sim
80$ \kms), which is
sufficient to include 93\% of lines in metal systems identified by
Bahcall et al. (1993).  The weighted mean $z_{ave}$ is defined as
\begin{displaymath}
z_{ave} \equiv {{\sum\displaylimits_i w_i^2 \; z_i} \over {\sum\displaylimits_i w_i^2}},
\end{displaymath}
where $w_i$ is the weight placed on the redshift of each line:
\begin{displaymath}
w_i = \cases{ {{W_{obs}} \over \sigma(W)} & $W_{obs} < 0.79 $ FWHM \cr
	      {{0.79 FWHM} \over \sigma(W)} & $W_{obs} \geq 0.79 $ FWHM},
\end{displaymath}
where FWHM is the spectral resolution of each grating in \AA.
We define the equivalent width cut-off as the limit for unsaturated lines.
Once a line becomes saturated, the accuracy of the line center depends
only on the S/N of the spectrum.  Lines which exceed the saturation criterion
are marked in Table 4.

\item 
We calculate expected positions of the remaining lines in these system, and we
measure wavelengths and equivalent widths, or limits for these lines.

\item 
We search for additional identifications for the new $2\sigma$ lines
amongst all systems with at least two $5\sigma$ lines.

\item 
As a further constraint, we use the ratios 
\begin{displaymath}
R_{H~I} = {W(1215) \over W(1025)},
\end{displaymath}
and
\begin{displaymath}
R_{O~VI} = {W(1032) \over W(1038)}.
\end{displaymath}
If Ly$\beta$ (1025) or O~VI(1038)
is not present, then the ratios are calculated using 
the $2\sigma $ equivalent
width limits. A ratio is judged acceptable if it 
falls between $1 - \sigma (R)$ and $2 + \sigma (R)$ for O~VI,
and $1 -\sigma (R)$ and $5.27 + \sigma  (R)$ for H~I, where $\sigma  (R)$ is 
the error on the ratio obtained from the errors on the individual $W$
values, and the numerical values are for unsaturated and saturated lines.
We discard systems which do not have at least one acceptable ratio.
\end{enumerate}

We wrote a computer code to follow the above procedures, 
and we ran this code with both
true and false rest wavelengths. The false wavelengths
did not have the same ratios as the true wavelengths of any 
strong lines.  We then counted the number of O~VI systems identified with
both true and false wavelengths.
The above procedures were then adjusted to give a true/false ratio of 5:1.
We also ran tests with true rest wavelengths and random absorption line
positions, which gave the same result.
 
Unfortunately, we did not accept lone O~VI lines, because the number of false
identifications was similar to the true number.  
In figure 1, we show the number
Our simulations show 
that we must insist that all systems have $\geq 3$
lines to keep the expected number of false 
O~VI identifications $<$ 1 per spectrum.  
Simulations indicate that \Lya should be stronger than O~VI
in either collisional or photo- ionization (VTB94),
so we use lines of the Lyman series to confirm O~VI identifications. 
Although we exclude possible O~VI detections,
it is necessary to detect a third line to keep false identifications to
a minimum.
A Lyman line must be detected at the O~VI redshift with the
following stipulations:

\bigskip
\begin{enumerate}
\item 
If \Lya exists and $R(H)$ is acceptable, then the system is accepted.

\item 
If \Lyb exists without a \Lya, the system is discarded.

\item 
If \Lyb exists and has an acceptable equivalent width
ratio with \Lyg, and \Lya cannot be checked, the system is
accepted. 
\end{enumerate}
\bigskip
 
No method will throw out all false systems and keep all real ones.
For example, consider pairs of $5 \sigma $ lines which are identified
as Ly$\alpha$ and Ly$\beta$.  Simulations show that such pairs do occur 
commonly, even in HST spectra at low $z$:
we expect 2 false pairs of $5 \sigma $ lines in every QSO.
But in real data we see about 6 such Ly$\alpha$ -- Ly$\beta$ pairs per QSO, 
so 67\% are real. We look for $2 \sigma $ O~VI lines in each such system.

\section{FREQUENCY OF OXYGEN VI}

In Table 2, we list the number of O~VI systems
identified with three different sample
criteria.  In Sample A, the absorption system identified with O~VI must have at 
least three lines with a redshift within $\Delta z = 0.0005$ of
the weighted mean redshift of the system.  All lines must have a significance
level $\geq 2\sigma$. 
Sample B is a subset of A where at least two of the lines must have a 
significance level of $\geq 5\sigma$.
Sample C is a subset of A, with the constraint that both lines of the 
O~VI doublet must have a rest equivalent width $W_{rest}\geq $ 0.21 \AA,  
which is the minimum 2$\sigma$ rest equivalent width 
which could be seen in all the spectra.
Sample C is the most stringent and complete
sub-sample and will be used to measure the density of O~VI systems.

Also shown in the same table is the number
of O~VI identifications per object using false rest wavelengths.  
Simulations were run with
10 sets of false wavelengths for \Lya, \Lyb, \Lyb and O~VI. 
We used the same criteria as with the real wavelengths.
The average number of false identifications is shown in the table, 
and gives an estimate of the number of identifications in the true sample 
which are due to chance.

Samples A and B differ by less than we expected, because there is only
one system in A that does not satisfy the requirements of B. 
But A contains twice as many false IDs
as B, so B is a significantly cleaner sample.
The line density
is actually higher in B than A, but the difference is not significant.

With 11 objects in this preliminary survey, we can estimate the number of O~VI absorbers 
per unit redshift.  
\begin{displaymath}
N(z) = {{\rm Total~ True - Total~ False}\over{\Delta z_{tot}}}
\end{displaymath}
where
\begin{displaymath}
\sigma_N(z) = {\sqrt{\rm Total~ True + Total~ False} 
\over{\Delta z_{tot}}}
\end{displaymath}
where $\Delta z_{tot} =4.73$. 

Table 3 contains system number densities for a variety of absorption systems. 
The first three are our samples of O~VI, and the fourth is a sample
of \Lya--\Lyb pairs in which we found W(1025) $\geq$ 0.30\AA\
For comparison with other samples, we use Sample C, which is most complete.
At low $z$ systems with O~VI are as common as the other main types of system:
Mg~II, C~IV and LLS.

\subsection{Individual Systems with O~VI}
 
Table 4 shows the twelve systems of Sample A.
For each of these systems we searched for all other
strong lines which would appear in the observed wavelength range,
including: C~IV 1548, 1550, N~V 1238, 1242,
Si~III~1206, Si~II~1190, 1193, 1260,  Si~IV~1393, 1402, N~III~989, 
C~III~977, and C~II~1334.
Two of these systems show H~I and O~VI only,
but all the others show at least one other element, and 
have $\geq 5$ lines each.
Lines are accepted into a system if their redshift is within
$\pm$0.0005 of the weighted mean.

In Figure 2 we show the absorption line positions on the HST spectra.
Note that the \Lyb, O~VI(1032) and O~VI(1038) lines form a well resolved
uniformly spaced triplet, which helps visual identifications.  

For each of the 12 O~VI systems of Sample A,
Figure 3 shows enlarged plots of \Lya, \Lyb, O~VI and
C~IV all on a velocity scale given by the listed mean redshift.
We do not show lines which are below $2\sigma$ significance.
Line widths are all set by the FOS resolution.
Comments on individual systems are in the appendix.

\subsection{Contamination of O~VI(1038)}

Lu \& Savage (1993) point out that O~VI 1037.62 can be contaminated by
C~II 1036.34 and O~I 1039.23, especially in low ionization systems.
 
Lines O~VI(1038) and C~II(1036) are separated by
$\Delta \lambda = 1.32 $ so to resolve these
lines we need spectra with resolution $R \geq$ 785.  The
FOS high resolution spectra have R $\approx$ 1300 so 
we would expect, assuming usual $b$ parameters $\leq$ 75 km~s$^{-1}$,
that these lines should be resolved in this survey.  
We do just resolve these lines, and we fitted each to get their individual
equivalent widths which are listed in Table 5.

As a check we measured lines C~II 1334.53 and O~I 1302.17 which should
both be stronger than the contaminants C~II 1036.34 and O~I 1039.23
because their $\lambda f$ values are larger by 1.3 and 6.6 times respectively.
We follow this procedure:
\begin{enumerate}
\item Measure width of C~II(1334).
\item Calculate line width of C~II(1036) with ratio of oscillator strengths,
assuming lines are linear.
\item Measure total width from 1036--1038 \AA.
\item Subtract calculated C~II(1036) from the combined width of step 3
\end{enumerate}

In Table 5, we list the equivalent widths of O~VI(1032), O~VI (1036-1038)
and C~II(1334). The O~VI 1036-1038
widths in Table 5 includes both O~VI and C~II, whereas those in Table 4 
are for O~VI alone. We detect C~II(1334) in four O~VI
systems,
and for each of these systems, we subtract
W(C~II,1036) from $W_r$(1036-1038) to calculate the "subtracted" 
doublet ratio.
If absorption is due to O~VI and C~II alone, this would provide the correct
width of O~VI(1038).  For all the systems of Sample A,
both the fitted and subtracted doublet ratios fall within the allowed 1$\sigma$ errors.
The fitting procedure contains less steps and is subject to less
uncertainty than the subtracting precedure.  Since the spectra of this
survey have adequate resolution to separate C~II(1036) and O~VI(1038),
we conclude that the measured widths and ratios
obtained by fitting O~VI
and C~II (Table 4) independently are more accurate.

The other possible contamination due to O~I is ruled out.  O~I 
$\lambda1302$ was not detected in any systems associated with O~VI.
So the contribution from the weaker O~I(1039) is safely
assumed negligible.

\subsection{Column Densities and Velocity Dispersions}

We employ the doublet ratio method to estimate column densities 
and velocity dispersions where possible.
Table 6 shows our calculations for O~VI, C~IV, and H~I in systems
where two or more ions are present.  
We calculate the 1$\sigma$ errors in Table 6 by allowing 
the equivalent widths to assume any value with their 1$\sigma$
errors and taking the extreme results as limits.
This is an over estimation of the random errors which partially
compensates for the systematic errors of measuring equivalent
widths of unresolved lines in the \Lya forest.  Wavelengths and
f-values used in the column density calculations were taken from
Morton (1991).

As seen from Table 6, our search was only sensitive to 
column densities log(N) $>$ 14.4.  This is consistent with the minimum
equivalent width threshold, $W_r = 0.21$ \AA.
Figure 4
shows the allowable column densities of a single component for a given
doublet ratio.  Our survey of O~VI is not sensitive to {\it linear} single
component absorbers.  
High velocity dispersion ($>$ 100 \kms) is strong evidence for multiple 
absorbers in the system.  

\subsection{Collisional or Photo- ionization?}

HST FOS spectra do not have enough spectral resolution to give 
accurate column densities and velocity dispersions,
so it is difficult to determine if the gas is photoionized or 
collisionally ionized.  
So we use the relative strengths of the O~VI, C~IV, and \Lyb lines
to indicate the level of ionization.
We define two equivalent width ratios:
\begin{displaymath}
R_{OB} = {O~VI(1032) \over Ly\beta}, 
\end{displaymath}
and
\begin{displaymath}
R_{OC} = {O~VI(1032) \over C~IV(1548)}.
\end{displaymath}
In Table 7 we give values for these ratios. \Lyb
is seen in 11 systems, and C~IV in 8.

Seven out of the eight systems in which C~IV and O~VI are detected 
have similar ratios:
$R_{OC} <$ 1.2 and $R_{OB} <$ 1 corresponding to medium ionization.  
Although C~IV and O~VI are both detected, 
C~IV is the dominant ion and the system should be labeled as such.  
Five of the seven medium ionization systems exhibit multiple 
low ionization lines (i.e. C~II, C~III, N~III, Si~II, Si~III).  
The ionization energies in each of the five systems range from ~11ev to 114eV. 
Such a wide range of ionizations, and therefore a large variety of ions, 
cannot be modeled with pure collisional ionization in a single simple cloud.  
As was restated by Verner and Yakolev (1990), the large
variety of ions found in absorption systems is strong evidence that power
law ionizing radiation
is present.
The eighth system, at $z_{abs} = 1.08$ of PG~1206+4557, has 
$R_{OC} =$ 4.17 and $R_{OB} =$ 2.02.  This system has significantly
different values and is classified as high ionization.   
Only collisional ionization at high temperatures, $T > 10^5 K$, or 
extreme photoionization, $\chi > 0.1$, can produce the measured ratios 
($\chi$ is the ratio
of the number of photons with energies above the Lyman edge to the
number of particles). 
Since C~IV is so weak and N~V is undetected, 
we believe that the system is collisionally ionized.

In the remaining four systems in which O~VI was detected, 
none had a corresponding C~IV
detection.  Each was assigned an upper limit $2\sigma$ 
equivalent width associated with
the non-detection of C~IV.  These systems are marked on Figure 5 
with arrows starting at the lower limit of the ratio.  In the
system $z_{abs} = 0.746$ of PKS~1424-1120, neither C~IV nor
\Lyb are detected and a lower limit is assigned for each ratio.
These systems are labeled as high ionization since O~VI
is the dominant metallic ion.  

\subsection{Cosmological Mass Density}

The cosmological mass density $\Omega_{OVI}(z)$ of O~VI is defined 
as the comoving mass density of O~VI in terms of the current critical
density.

\begin{displaymath}
\Omega_{OVI} = {m_{OVI}\over{\rho_c\,c\,H_o^{-1}}}
        {{\sum\displaylimits_{i} N_i(OVI)} \over
        {{\sum\displaylimits_{i} \Delta X_i}}},
\end{displaymath}
where $m_{OVI}$ is mass of the Oxygen ion, $\rho_c$ is the current critical
density, $H_o$ is the current Hubble parameter,
$N_i(OVI)$ is total O~VI column density towards the $i$th QSO over an 
absorption path distance $\Delta X_i$ defined as

\begin{displaymath}
\Delta X = \cases{{1 \over 2} \{[(1+z_{max})^2-1] - [(1+z_{min})^2-1]\}  
& ($q_0$ = 0), \cr
                    {2 \over 3} \{[(1+z_{max})^{3/2}-1] - [(1+z_{min})^{3/2}-1]\} 
& ($q_0$ = 0.5). }
\end{displaymath}

Using values of the redshift range in Table 2, 

\begin{displaymath}
{\sum\displaylimits_{i} \Delta X_i} = \cases{9.24 & ($q_0$ = 0), \cr
                    6.60 & ($q_0$ = 0.5). } 
\end{displaymath}

The standard error of the mass density is estimated as

\begin{displaymath}
\sigma_\Omega = {1 \over {(1- {1 \over m})^{1/2}}}
        {{\sqrt{\sum\displaylimits_i
        [N_i(OVI) - \langle N(OVI) \rangle]^2)}} \over { \Delta X_{tot} }},
\end{displaymath}
where m is the total number of absorption systems in the sample, 
and $\langle N(OVI) \rangle$
is the mean column density of the sample.

Lower limits of the column density in each absorption system are calculated
with the assumption that the O~VI absorption features are unsaturated.
The column density of the absorber is then proportional to the rest
equivalent width (Spitzer 1978) 

\begin{displaymath}
N(cm^{-2}) = {{1.13 \times 10^{20} \; W(\AA)} \over {\lambda^2(\AA) \, F_{ij}}},
\end{displaymath}
where $F_{ij}$ is the oscillator strength of the transition
$\lambda$ is the rest wavelength of the transition
W is the rest equivalent width.  The absorption features of
O~VI $\lambda$1038 are weaker and less saturated.
Using the six systems of Sample C and the rest equivalent width of
O~VI(1038), we obtain

\begin{displaymath}
{\sum\displaylimits_{i} N_i(OVI)} \geq 5.9 \; \times \; 10^{15} cm^{-2}.
\end{displaymath}

We arrive at an estimate for the cosmological mass density of O~VI at 
$z_{ave} = 0.9$, 

\begin{displaymath}
\Omega_{OVI} \geq \cases{1.4 \pm 0.4 \times 10^{-8} h^{-1} & ($q_0$ = 0), \cr
                           2.0 \pm 0.6  \times
                                10^{-8} h^{-1} & ($q_0$ = 0.5), }
\end{displaymath}

where $H_o = 100 h\; km \; s^{-1} \; Mpc^{-1}$.

\subsection{Limit on Cosmic Metallicity}

It is now straightforward to place a lower limit on 
the mean cosmological metallicity at 
$z_{ave} = 0.9$.  If the cosmic metallicity is too low,
all the baryons in the universe are needed to account for
the O~VI seen in this survey.
Let $\zeta$(z) be defined as the ratio of
mean metallicity at redshift $z$
to the solar metallicity,

\begin{displaymath}
\zeta (z) = \mu \; \left(H \over O \right)_{Solar} \; 
                \left(O \over {OVI} \right) \;
                 \; {{\Omega_{OVI}(z)} \over {\Omega_b}}
\end{displaymath}
where $\mu$ is the mean molecular weight, which equals 1.3 for a mixture
of 25\% Helium and 75\% Hydrogen by mass; $(H/O)_{Solar}
 = 1174.9$ (\cite{and89}); and $\Omega_b$ is the cosmological baryon
mass density.
The ionization fraction O/O~VI is impossible to calculate for each absorption
system with the current data, but a lower limit is placed on the metallicity
by assuming O/O~VI = 1.  A recent measurement of deuterium at high redshift by
(Tytler \& Fan 1994) yields $\Omega_b^{-1}(z) = 0.023 h^{-2}$.

\begin{displaymath}
\zeta (z) \geq \cases{8.9 \pm 2.7 \times 10^{-4} h & ($q_0$ = 0), \cr
                        1.4 \pm 0.4 \times 10^{-3} h & ($q_0$ = 0.5). }
\end{displaymath}

The calculation is subject to small number statistics, and we can estimate
the sensitivity of the lower limit by removing one system of Sample C.
If we discard the system with the largest equivalent width 
(\zabs=0.927 towards 1206+4557), the lower limit is reduced by 35\%.
The calculated lower limit is a safe estimate for the following 
reasons: We have assumed
that all Oxygen at $z_{ave} = 0.9$ is in the form of O~VI, that all absorption
is linear, and that we have detected all O~VI along the absorption path.  

\section{SUMMARY}

O~VI absorption is as common at $z_{ave} = 0.9$ as C~IV
and Mg~II absorption.  Using a minimum rest
equivalent width of $W_r = 0.21$ \AA~ and 
accounting for the average number of chance
coincidences per spectra, the number density of 
O~VI absorbers per redshift is 

\begin{displaymath}
\langle N_{OVI}(z=0.9) \rangle = 1.0 \pm 0.6 \; .
\end{displaymath}

Using a systematic computer search, we found 12 redshifts with
$\geq 2\sigma$ equivalent widths at O~VI wavelengths.  
The O~VI systems were categorized as either medium or high ionization,
Seven systems are medium ionization, with 5 showing multiple
low ionization lines.  The 5 high ionization systems have O~VI stronger
than both C~IV and \Lyb .  

We summed over all O~VI absorption with widths above the threshold
$W_r > 0.21$ \AA~to estimate the cosmological mass density at $z = 0.9$.
Since we could not resolve velocity structure, we calculated a lower limit
by assuming linear absorption, 

\begin{displaymath}
\Omega_{OVI} \geq \cases{1.4 \pm 0.4 \times 10^{-8} h^{-1} & ($q_0$ = 0), \cr
                           2.0 \pm 0.6  \times
                                10^{-8} h^{-1} & ($q_0$ = 0.5). }
\end{displaymath}
We proceed further by assuming
all Oxygen is in the form of O~VI and place a lower limit on the metallicity
in solar units at $z=0.9$, 

\begin{displaymath}
\zeta (z) \geq \cases{8.9 \pm 2.7 \times 10^{-4} h & ($q_0$ = 0), \cr
                        1.4 \pm 0.4 \times 10^{-3} h & ($q_0$ = 0.5). }
\end{displaymath}

The survey for hot gas, such as O~VI, must be pursued in the future.  In addition 
to better statistics at low redshift, the survey should be extended to high redshift
to measure changes in the abundance of hot gas.  Ions with
higher energies, such as Ne~VIII and Mg~X, should be studied.  
The launch of the Hubble Space Telescope has allowed us to quantify the amount
of 100 eV intergalactic gas for the first time. 

\acknowledgments
We are grateful to Chris Davis for assisting in the reduction of the
archived HST spectra.  

\clearpage


\begin{table*}
\begin{center}
\begin{tabular}{ccccccc}
QSO(1950) & Name & $z_{em}$  & V & FOS Grating &
Offset(\AA) & Exposure (s) \cr
\tableline
0122-0021 & PKS~0122-00 & 1.070 & 16.70 & 190H  & +1.3 & 3000 \cr
	  &		&	&	& 270H	& +0.8 & 720 \cr
0454-2203 & PKS~0454-22 & 0.534 & 16.1  & 130H  & 0.0  & 6500 \cr
	  &	 	& 	&	& 190H  & 0.0  & 5900 \cr
	  &		&	&	& 270H	& 0.0  & 2000 \cr
1206+4557 & PG~1206+459 & 1.158 & 15.79 & 190H  & +0.8 & 4500 \cr
	  &		& 	& 	& 270H  & +0.9 & 1000 \cr
1317+2743 & Ton~153 	& 1.022	& 15.98	& 190H	& +1.5 & 4500 \cr
	  &		&	&	& 270H  & +1.4 & 1000 \cr
1352+0106 & PG~1352+011 & 1.121	& 16.03 & 190H  & +2.3 & 5300 \cr
	  &		&	&	& 270H  & +1.6 & 1150 \cr
1407+2632 & PG~1407+265 & 0.944 & 15.73 & 190H  & +1.5 & 4250 \cr
	  &		& 	&	& 270H  & +1.9 & 1125 \cr
1424-1150 & PKS~1424-11 & 0.806 & 16.49 & 190H  & +1.5 & 6600 \cr
	  &		&	&	& 270H  & +1.7 & 1900 \cr
1435+6349 & S4-1435+63  & 2.068 & 15.0  & 270H  & 0.0  & 3200 \cr 
1522+1009 & PG~1522+101 & 1.321 & 15.74 & 190H  & +0.6 & 9500 \cr
	  &		&	&	& 270H  & +1.9 & 2650 \cr
1634+7037 & PG~1634+706 & 1.334 & 14.90 & 190H  & +0.9 & 5300 \cr
	  &		&	&	& 270H  & +0.7 & 6750 \cr
2340-0340 & PKS~2340-03 & 0.896 & 16.02 & 190H  & +0.2 & 4100 \cr
\end{tabular}
\end{center}

\tablenum{1}
\caption{List of Quasar Spectra} 

\end{table*}

\clearpage

\begin{table*}
\begin{center}
\begin{tabular}{ccccccccccc}
{Object} &
\multispan{2} \hfil {Lyman-$\alpha,\beta$} \hfil &
\multispan{6} \hfil {O~VI(1032,1038)} \hfil & Redshift & Minimum \cr
 & True ID & False ID &
\multispan{3} \hfil {True ID} \hfil &
\multispan{3} \hfil {False ID} \hfil & Range & Observed \cr
 &         &          & A & B & C & A & B & C & ${(\Delta z)}$ & {Width(\AA)} \cr
\tableline
0122-0021 & 5 & 2.2 & 2 & 2 & 2 & 0.2 & 0.1 & 0.0 & 0.61--1.07 & 0.31 \cr
0454-2203 & 4 & 0.7 & 1 & 1 & 1 & 0.4 & 0.1 & 0.1 & 0.31--0.53 & 0.28 \cr
1206+4557 & 6 & 1.1 & 3 & 2 & 2 & 0.2 & 0.2 & 0.1 & 0.61--1.16 & 0.24 \cr
1317+2743 & 2 & 0.6 & 0 & 0 & 0 & 0.3 & 0.2 & 0.1 & 0.61--1.02 & 0.21 \cr
1352+0106 & 5 & 1.1 & 1 & 1 & 1 & 0.3 & 0.3 & 0.2 & 0.61--1.12 & 0.23 \cr
1407+2632 & 6 & 0.3 & 0 & 0 & 0 & 0.1 & 0.1 & 0.1 & 0.61--0.94 & 0.31 \cr
1424-1150 & 2 & 0.1 & 2 & 2 & 0 & 0.0 & 0.0 & 0.0 & 0.61--0.81 & 0.29 \cr
1435+6349 & 0 & 0.6 & 0 & 0 & 0 & 0.7 & 0.2 & 0.2 & 1.61--2.08 & 0.23 \cr
1522+1009 & 2 & 0.6 & 0 & 0 & 0 & 0.7 & 0.4 & 0.3 & 0.61--1.32 & 0.16 \cr
1634+7037 & 1 & 0.2 & 2 & 2 & 0 & 0.4 & 0.1 & 0.1 & 0.75--1.33 & 0.17 \cr
2340-0340 & 3 & 0.4 & 1 & 1 & 0 & 0.2 & 0.1 & 0.1 & 0.61--0.90 & 0.25 \cr
\tableline
\noalign{\vskip 3pt}
Total      & 39 & 7.9 & 12 & 11 & 6 & 3.5 & 1.8 & 1.3 & 4.73 & \cr
\tablenotetext{}
{A: System must have at least three lines with a redshift within $\Delta z
= 0.0005$ of
the mean redshift of the system.  All lines must be
$\geq 2\sigma$.}
 
\tablenotetext{}{B: Subset of A where at least two lines are $\geq 5\sigma$.}
 
\tablenotetext{}{C: Subset of B where both
O~VI lines have a rest equivalent width $\geq 0.21 $ \AA.}

\end{tabular}
\end{center}

\tablenum{2}
\caption{Frequency Of Absorption Systems with O~VI Doublets}
\end{table*}

\clearpage

\begin{table*}
\begin{center}
\begin{tabular}{rrccccc}
Ion & Sample & True & False & $\overline z$ & N(z) & ref \cr
\tableline
\noalign{\vskip 5pt}
 & & \multispan{2} Low~Redshift & & \cr
\noalign{\vskip 5pt}
O~VI & A        & 12 & 3.5 & 0.9 & $1.8 \pm{0.8}$ & 1 \cr
    & B & 11 & 1.8 & 0.9 & $1.9 \pm{0.7}$ & 1 \cr
    & C & 6  & 1.3 & 0.9 & $1.0 \pm{0.6}$ & 1 \cr
HI  & \Lya-\Lyb & 39 & 8 & 0.9 & $6.5 \pm{1.4}$ & 1 \cr
HI  & \Lya & \multispan 2 \hfil {W(1215) $>$ 0.32\AA} 
	\hfil & 0.7 & $21.8 \pm{2.1}$ & 2 \cr 
HI  & LLS & \multispan 2 \hfil {$\tau > 0.4$} 
	\hfil & 0.9 & $1.1 \pm{0.3}$ & 2 \cr
C~IV & \multispan 3 \hfil {W(1548) $>$ 0.30\AA} 
	\hfil & 0.3 & $0.87 \pm{0.43}$ & 2 \cr
Mg~II & \multispan 3 \hfil {W(2796) $>$ 0.60\AA} 
	\hfil & 0.5 & $0.26 \pm{0.08}$ & 3 \cr
Mg~II & \multispan 3 \hfil {W(2796) $>$ 0.30\AA}
	\hfil & 0.9 & $1.0 \pm{0.25}$ & 4 \cr
\tablerefs{
(1) This Paper; (2) \cite{bac93}; (3) \cite{tyt87};
(4) \cite{fan95}.} 

\end{tabular}
\end{center}

\tablenum{3}
\caption{Number Density Of Various Absorption Systems}

\end{table*}

\clearpage

\small
\begin{planotable}{clccccc}
\tablewidth{0pt}
\tablenum{4}
\tablecaption{Oxygen~VI System Line List\tablenotemark{a}}
\tablehead{
\colhead{Wavelength} & \colhead{ID} & \colhead{$z_{obs}$} & \colhead{$W_{obs}$}
& \colhead{$\sigma(W)$} & \colhead{SL\tablenotemark{b}} 
& \colhead{LR\tablenotemark{c}}  \nl
\colhead{(\ang)} & & & \colhead{(\ang)} & \colhead{(\ang)}}
\startdata
 & \multispan{2} \hfil {0122-0021} \hfil
 &  \multispan{2} \hfil  \zem = 1.070  \hfil
 & V=16.1\tablenotemark{d} & \nl
\noalign{\vskip 3pt}
\tableline
\noalign{\vskip 3pt}
\multispan 1 {1} \hfil & \multispan 5 \hfil $z_{ave}$ = 0.95336, 
		$\sigma_z$ = 0.00031 A,B,C\tablenotemark{e} \hfil & \nl
\noalign{\vskip 3pt}
\tableline 
\noalign{\vskip 3pt}
2374.84 &\Lya        &0.95352 &2.368 &0.205\tablenotemark{f} &11.55    &1.73  \nl
2003.88 &\Lyb        &0.95363 &1.367 &0.102\tablenotemark{f} &13.38    &2.19 \nl1900.05 &\Lyg        &0.95370 &0.623 &0.122     & 5.11& \nl
2015.31 &OVI(1032)   &0.95295 &1.211\tablenotemark{g} &0.098\tablenotemark{f} &12.31    &1.41 \nl
2026.92 &OVI(1038)   &0.95343 &0.861 &0.095     & 9.03 & \nl
1908.48 &CIII(977)   &0.95337 &1.395 &0.118\tablenotemark{f} &11.82 & \nl
1933.90 &NIII(989)   &0.95383 &0.694 &0.102     & 6.80 & \nl
2356.67 &SiIII(1206) &0.95331 &0.467 &0.200     & 2.35 & \nl
        &NV(1238)    &        &\lq0.410\tablenotemark{h} & & & \nl
        &NV(1242)    &        &\lq0.410\tablenotemark{h} & & & \nl
2606.32 &CII(1334)   &0.95298 &0.360 &0.154     & 2.33 & \nl
3024.09 &CIV(1548)   &0.95329 &1.989 &0.180\tablenotemark{f} &11.05    &1.13 \nl3029.12 &CIV(1550)   &0.95330 &1.763 &0.180\tablenotemark{f} & 9.79 & \nl
\noalign{\vskip 3pt}
\tableline
\noalign{\vskip 3pt}
\multispan 1 {2} \hfil & \multispan{5} \hfil { $z_{ave}$ = 0.96686, 
		$\sigma_z$ = 0.00031 A,B,C\tablenotemark{e}} \hfil & \nl
\noalign{\vskip 3pt}
\tableline
\noalign{\vskip 3pt}
2391.13 &\Lya        &0.96693 &1.873 &0.202\tablenotemark{f} & 9.29    &2.33 \nl2017.20 &\Lyb        &0.96662 &0.804 &0.098     & 8.22    &0.58\tablenotemark{i} \nl
1912.64 &\Lyg        &0.96664 &1.394 &0.116\tablenotemark{f} &12.04 & \nl
2030.11 &OVI(1032)   &0.96730 &0.708 &0.095     & 7.47    &1.45 \nl
2041.14 &OVI(1038)   &0.96714 &0.488 &0.093     & 5.25 & \nl
        &NV(1238)    &        &\lq0.410\tablenotemark{h} & & & \nl
        &NV(1242)    &        &\lq0.410\tablenotemark{h} & & & \nl
3044.48 &CIV(1548)   &0.96646 &0.629 &0.180     & 3.49 &\gq1.74 \nl
	&CIV(1550)   &        &\lq0.36\tablenotemark{h} & & & \nl
\noalign{\vskip 3pt}
\tableline
\noalign{\vskip 3pt}
 & \multispan{2} \hfil {0454-2203} \hfil
 &  \multispan{2} \hfil { \zem = 0.534} \hfil
 & V=16.1\tablenotemark{d} & \nl
\noalign{\vskip 3pt}
\tableline
\noalign{\vskip 3pt}
\multispan 1 {3} \hfil & \multispan{5} \hfil { $z_{ave}$ = 0.41261, 
		$\sigma_z$ = 0.00025 A,B,C\tablenotemark{e}} \hfil & \nl
\noalign{\vskip 3pt}
\tableline
\noalign{\vskip 3pt}
1717.28 &\Lya        &0.41262 &0.489 &0.036     &13.58    &0.72\tablenotemark{i} \nl
1448.75 &\Lyb        &0.41242 &0.683 &0.060     &11.46 &\gq5.85\tablenotemark{i} \nl
1457.61 &OVI(1032)   &0.41251 &0.851 &0.059\tablenotemark{f} &14.38    &1.40 \nl1466.13 &OVI(1038)   &0.41298 &0.608 &0.059     &10.34    & \nl
1380.00 &CIII(977)   &0.41246 &0.691 &0.060     &11.51    & \nl
1397.89 &NIII(989)   &0.41230 &0.153 &0.060     & 2.55    & \nl
1704.29 &SiIII(1206) &0.41259 &0.809 &0.040     &20.23    & \nl
1750.49 &NV(1238)    &0.41303 &0.295 &0.030     & 9.83    & \nl
        &NV(1242)    &        &\lq0.06\tablenotemark{h} & & & \nl
	&CIV(1548)   &        &\lq0.18\tablenotemark{h} & & & \nl
	&CIV(1550)   &        &\lq0.18\tablenotemark{h} & & & \nl
\tablebreak
 & \multispan{2} \hfil {1206+4557} \hfil
 &  \multispan{2} \hfil { \zem = 1.158} \hfil
 & V=15.79\tablenotemark{d} & \nl
\noalign{\vskip 3pt}
\tableline
\noalign{\vskip 3pt}
\multispan 1 {4} \hfil & \multispan{5} \hfil { $z_{ave}$ = 0.73385, 
		$\sigma_z$ = 0.00032 A\tablenotemark{e}} \hfil & \nl
\noalign{\vskip 3pt}
\tableline
\noalign{\vskip 3pt}
2107.78 &\Lya        &0.73384 &0.847 &0.051     &16.54    &0.75 \nl
1779.34 &\Lyb        &0.73472 &1.131\tablenotemark{i} &0.081     &13.96 & \nl
1789.01 &OVI(1032)   &0.73365 &0.279 &0.077     & 3.62    &1.15 \nl
1799.50 &OVI(1038)   &0.73426 &0.243 &0.076     & 3.19 \nl
        &NV(1238)    &        &\lq0.44\tablenotemark{h} & & & \nl
        &NV(1242)    &        &\lq0.44\tablenotemark{h} & & & \nl
2684.95 &CIV(1548)   &0.73424 &0.371 &0.066     & 5.62 \nl
	&CIV(1550)   &        &\lq0.13\tablenotemark{h} & & & \nl
\noalign{\vskip 3pt}
\tableline
\noalign{\vskip 3pt}
\multispan 1 {5} \hfil & \multispan{5} \hfil {$z_{ave}$ = 0.92703, 
		$\sigma_z$ = 0.00042 A,B,C\tablenotemark{e}} \hfil & \nl
\noalign{\vskip 3pt}
\tableline
\noalign{\vskip 3pt}
2342.55 &\Lya        &0.92696 &4.490 &0.122\tablenotemark{f} &36.80    &1.50 \nl1976.11 &\Lyb        &0.92656 &3.001 &0.068\tablenotemark{f} &43.87    &2.27 \nl1874.54 &\Lyg        &0.92747 &1.320 &0.076\tablenotemark{f} &17.32 & \nl
1987.73 &OVI(1032)   &0.92623 &2.328 &0.067     &34.75    &0.94 \nl
1999.29 &OVI(1038)   &0.92680 &2.490 &0.067     &37.16 & \nl
1881.66 &CIII(977)   &0.92592 &4.066\tablenotemark{j} &0.076     &53.50    & \nl1907.79 &NIII(989)   &0.92745 &1.113 &0.078     &14.27 & \nl
2294.61 &SiII(1190)  &0.92757 &0.295 &0.046     & 6.41 & \nl
2300.26 &SiII(1193)  &0.92766 &0.488 &0.046     &10.61 & \nl
2326.10 &SiIII(1206) &0.92797 &1.379 &0.124     &11.12 & \nl
2386.94 &NV(1238)    &0.92679 &1.792 &0.117     &15.32    &1.64 \nl
2394.23 &NV(1242)    &0.92648 &1.093 &0.117     & 9.34 & \nl
2573.14 &CII(1334)   &0.92812 &1.835 &0.075     &24.47 & \nl
2686.56 &SiIV(1393)  &0.92757 &1.294 &0.067     &19.31    &1.20 \nl
2703.75 &SiIV(1402)  &0.92743 &1.079 &0.067     &16.10 & \nl
2984.04 &CIV(1548)   &0.92742 &4.440 &0.096     &46.25    &1.62 \nl
2988.88 &CIV(1550)   &0.92735 &2.737 &0.096     &28.51 & \nl
\noalign{\vskip 3pt}
\tableline
\noalign{\vskip 3pt}
\multispan 1 {6} \hfil & \multispan{5} \hfil { $z_{ave}$ = 1.08275, 	
		$\sigma_z$ = 0.00130 A,B,C\tablenotemark{e}} \hfil & \nl
\noalign{\vskip 3pt}
\tableline
\noalign{\vskip 3pt}
2529.96 &\Lya        &1.08113 &1.243 &0.105     &11.84    &0.98 \nl
2134.35 &\Lyb        &1.08298 &1.262 &0.046\tablenotemark{f} &27.43    &2.69 \nl2023.47 &\Lyg        &1.08060 &0.469 &0.057     & 8.23 & \nl
2149.35 &OVI(1032)   &1.08284 &2.543 &0.044\tablenotemark{f} &58.06    &1.18 \nl2161.21 &OVI(1038)   &1.08285 &2.160 &0.044\tablenotemark{f} &49.54 & \nl
2510.57 &SiIII(1206) &1.08087 &1.038 &0.112     & 9.27 & \nl
        &NV(1238)    &        &\lq0.17\tablenotemark{h} & & & \nl
        &NV(1242)    &        &\lq0.17\tablenotemark{h} & & & \nl
3224.76 &CIV(1548)   &1.08291 &0.605 &0.101     & 5.98    &1.20 \nl
3229.31 &CIV(1550)   &1.08239 &0.505 &0.101     & 5.00 & \nl
\tablebreak
 & \multispan{2} \hfil {1317+2743} \hfil
 &  \multispan{2} \hfil { \zem = 1.022} \hfil
 & V=15.98\tablenotemark{d} & \nl
\noalign{\vskip 3pt}
\tableline
\noalign{\vskip 3pt}
\multispan{7} \hfil {No O~VI Detected} \hfil \nl
\noalign{\vskip 3pt}
\tableline
\noalign{\vskip 3pt}
 & \multispan{2} \hfil {1352+0106} \hfil
 &  \multispan{2} \hfil { \zem = 1.121} \hfil
 & V=16.03\tablenotemark{d} & \nl
\noalign{\vskip 3pt}
\tableline
\noalign{\vskip 3pt}
\multispan 1 {7} \hfil & \multispan{5} \hfil { $z_{ave}$ = 0.66752, 
		$\sigma_z$ = 0.00034 A,B,C\tablenotemark{e}} \hfil & \nl
\noalign{\vskip 3pt}
\tableline
\noalign{\vskip 3pt}
2026.88 &\Lya        &0.66730 &2.920 &0.065\tablenotemark{f} &45.06    &1.34 \nl1710.59 &\Lyb        &0.66769 &2.177 &0.086\tablenotemark{f} &25.20  & \nl
1721.07 &OVI(1032)   &0.66782 &1.089 &0.084     &12.90    &2.46 \nl
1730.69 &OVI(1038)   &0.66794 &0.442 &0.083     & 5.35  & \nl
1985.62 &SiII(1190)  &0.66800 &0.503 &0.064     & 7.86  & \nl
1989.50 &SIII(1193)  &0.66724 &1.084 &0.064     &16.94  & \nl
2011.92 &SiIII(1206) &0.66757 &1.652 &0.065\tablenotemark{f} &25.41  & \nl
2066.08 &NV(1238)    &0.66778 &0.485 &0.053     & 9.15  &1.49 \nl
2072.16 &NV(1242)    &0.66733 &0.325 &0.052     & 6.25  & \nl
2225.39 &CII(1334)   &0.66754 &1.197 &0.047     &25.47  & \nl
2324.18 &SiIV(1393)  &0.66757 &1.225 &0.124     & 9.88  &0.73 \nl
2338.93 &SiIV(1402)  &0.66737 &1.681 &0.123     &13.67  & \nl
2581.02 &CIV(1548)   &0.66711 &2.219 &0.074     &29.98  &1.16 \nl
2585.92 &CIV(1550)   &0.66750 &1.921 &0.074     &25.96  & \nl
\noalign{\vskip 3pt}
\tableline
\noalign{\vskip 3pt}
 & \multispan{2} \hfil {1407+2632} \hfil 
 &  \multispan{2} \hfil { \zem = 0.944} \hfil 
 & V=15.73\tablenotemark{d} & \nl 
\noalign{\vskip 3pt}
\tableline
\noalign{\vskip 5pt}
\multispan{7} \hfil {No O~VI Detected} \hfil \nl
\noalign{\vskip 5pt}
\tableline
\noalign{\vskip 3pt}
 & \multispan{2} \hfil {1424-1150 }  \hfil
 &  \multispan{2} \hfil { \zem = 0.806 }  \hfil
 & V=16.49\tablenotemark{d} & \nl
\noalign{\vskip 3pt}
\tableline
\noalign{\vskip 3pt}
\multispan 1 {8} \hfil & \multispan{5} \hfil { $z_{ave}$ = 0.65516, 
		$\sigma_z$ = 0.00027 A,B\tablenotemark{e}} \hfil & \nl
\noalign{\vskip 3pt}
\tableline
\noalign{\vskip 3pt}
2012.11 &\Lya        &0.65514 &1.959 &0.083\tablenotemark{f} &23.72    &1.58 \nl1697.85 &\Lyb        &0.65528 &1.237 &0.101\tablenotemark{f} &12.22 & \nl
1707.88 &OVI(1032)   &0.65503 &0.313 &0.100     & 3.14    &0.77 \nl
1716.98 &OVI(1038)   &0.65473 &0.409 &0.098     & 4.16 & \nl
1997.45 &SiIII(1206) &0.65557 &0.299 &0.083     & 3.60 & \nl
        &NV(1238)    &        &\lq0.17\tablenotemark{h} & & & \nl
        &NV(1242)    &        &\lq0.17\tablenotemark{h} & & & \nl
2562.81 &CIV(1548)   &0.65535 &0.653 &0.137     & 4.77    &0.78 \nl
2567.06 &CIV(1550)   &0.65535 &0.842 &0.137     & 6.15 & \nl
\noalign{\vskip 3pt}
\tableline
\noalign{\vskip 3pt}
\multispan 1 {9} \hfil & \multispan{5} \hfil { $z_{ave}$ = 0.74654, 
		$\sigma_z$ = 0.00042 A,B\tablenotemark{e}} \hfil &  \nl
\noalign{\vskip 3pt}
\tableline
\noalign{\vskip 3pt}
2123.49 &\Lya        &0.74676 &0.463 &0.062     & 7.42 &\gq2.80 \nl
1801.81 &OVI(1032)   &0.74606 &0.542 &0.083     & 6.56    &1.91 \nl
1812.53 &OVI(1038)   &0.74682 &0.284 &0.081     & 3.51 \nl
        &NV(1238)    &        &\lq0.11\tablenotemark{h} & & & \nl
        &NV(1242)    &        &\lq0.11\tablenotemark{h} & & & \nl
	&CIV(1548)   &        &\lq0.22\tablenotemark{h} & & & \nl
	&CIV(1550)   &        &\lq0.22\tablenotemark{h} & & & \nl
\tablebreak
 & \multispan{2} \hfil {1435+6349} \hfil 
 &  \multispan{2} \hfil { \zem = 2.068} \hfil 
 & V=15.0\tablenotemark{d} & \nl 
\noalign{\vskip 3pt}
\tableline
\noalign{\vskip 5pt}
\multispan{7} \hfil {No O~VI Detected} \hfil \nl
\noalign{\vskip 5pt}
\tableline
\noalign{\vskip 3pt}
 & \multispan{2} \hfil {1522+1009} \hfil 
 &  \multispan{2} \hfil { \zem = 1.321} \hfil 
 & V=15.74\tablenotemark{d} & \nl 
\noalign{\vskip 3pt}
\tableline
\noalign{\vskip 5pt}
\multispan{7} \hfil {No O~VI Detected} \hfil \nl
\noalign{\vskip 5pt}
\tableline
\noalign{\vskip 3pt}
 & \multispan{2} \hfil {1634+7037} \hfil
 &  \multispan{2} \hfil { \zem = 1.334} \hfil
 & V=14.9\tablenotemark{d} & \nl
\noalign{\vskip 3pt}
\tableline
\noalign{\vskip 3pt}
\multispan 1 {10} \hfil & \multispan{5} \hfil { $z_{ave}$ = 1.04162, 
		$\sigma_z$ = 0.00017 A,B\tablenotemark{e} } \hfil &  \nl
\noalign{\vskip 3pt}
\tableline
\noalign{\vskip 3pt}
2481.92 &\Lya        &1.04161 &2.731 &0.037\tablenotemark{f} &74.21    &1.94 \nl2094.08 &\Lyb        &1.04157 &1.411 &0.052\tablenotemark{f} &27.34    &1.37 \nl1985.50 &\Lyg        &1.04156 &1.032 &0.068     &15.09    &1.05 \nl
1939.12 &\Lyd        &1.04174 &0.983 &0.073     &13.43    &1.25 \nl 
1914.87 &\Lye        &1.04187 &0.786 &0.078     &10.08 & \nl
2107.08 &OVI(1032)   &1.04188 &0.424 &0.049     & 8.58    &1.48 \nl
2118.31 &OVI(1038)   &1.04151 &0.287 &0.049     & 5.86 & \nl
1994.39 &CIII(977)   &1.04130 &0.406 &0.067     & 6.06 & \nl
2430.49 &SiII(1190)  &1.04172 &0.258 &0.067     & 3.85 & \nl
2436.10 &SiII(1193)  &1.04150 &1.532 &0.037     &41.85 & \nl
2463.84 &SiIII(1206) &1.04214 &1.121 &0.037     &30.29 & \nl
        &NV(1238)    &        &\lq0.07\tablenotemark{h} & & & \nl
        &NV(1242)    &        &\lq0.07\tablenotemark{h} & & & \nl
2724.81 &CII(1334)   &1.04178 &0.216 &0.031     & 6.92 & \nl
3160.21 &CIV(1548)   &1.04122 &1.044 &0.036     &28.83    &1.41 \nl 
3165.87 &CIV(1550)   &1.04148 &0.743 &0.036     &20.52 & \nl
\multispan 1 {11} \hfil & \multispan{5} \hfil { $z_{ave}$ = 1.14121, 
		$\sigma_z$ = 0.00028  A,B\tablenotemark{e} } \hfil & \nl
\noalign{\vskip 3pt}
\tableline
\noalign{\vskip 3pt}
2602.99 &\Lya        &1.14120 &1.088 &0.033     &32.52    &6.26 \nl
2196.12 &\Lyb        &1.14105 &0.174 &0.048     & 3.65 &\gq1.61 \nl
2209.94 &OVI(1032)   &1.14156 &0.242 &0.047     & 5.19    &1.20 \nl
2222.08 &OVI(1038)   &1.14152 &0.202 &0.046     & 4.42    & \nl
        &NV(1238)    &        &\lq0.07\tablenotemark{h} & & & \nl
        &NV(1242)    &        &\lq0.07\tablenotemark{h} & & & \nl
2698.25 &SiII(1260)  &1.14075 &0.533 &0.032     &16.66    & \nl
\tablebreak
 & \multispan{2} \hfil {2340-0340} \hfil
 &  \multispan{2} \hfil { \zem = 0.896} \hfil
 & V=16.02\tablenotemark{d} & \nl
\noalign{\vskip 3pt}
\tableline
\noalign{\vskip 3pt}
\multispan{1} {12} \hfil & \multispan{5} \hfil { $z_{ave}$ = 0.68416, 
		$\sigma_z$ = 0.00029 A,B\tablenotemark{e} } \hfil &  \nl
\noalign{\vskip 3pt}
\tableline
\noalign{\vskip 3pt}
2047.22 &\Lya        &0.68403 &1.764 &0.065 &26.97    &0.94\tablenotemark{i} \nl1727.90 &\Lyb        &0.68457 &1.882 &0.094 &19.94    &5.88\tablenotemark{i} \nl1637.57 &\Lyg        &0.68381 &0.320 &0.134 & 2.39 & \nl
1737.92 &OVI(1032)   &0.68415 &0.525 &0.093 & 5.62    &1.68 \nl
1747.65 &OVI(1038)   &0.68428 &0.312 &0.092 & 3.39 & \nl
1645.67 &CIII(977)   &0.68438 &1.657 &0.127 &13.05 & \nl
2031.65 &SiIII(1206) &0.68392 &0.466 &0.067 & 6.96 & \nl
2086.02 &NV(1238)    &0.68388 &0.148 &0.062 & 2.39 &\gq1.19 \nl
        &NV(1242)    &        &\lq0.12\tablenotemark{h} & & & \nl
2247.02 &CII(1334)   &0.68375 &0.201 &0.044 & 4.57 & \nl
\tablenotetext{a}{ Table contains 11 Objects, 20 Spectra, \& 12 O~VI detections.}
\tablenotetext{b}{ Significance Level: $W_{obs} / \sigma(W)$.}
\tablenotetext{c}{ Line Ratio:  Ratio of equivalent widths is calculated for the 
	following pairs of lines.  
	(\Lya/\Lyb, \Lyb/\Lyg, O~VI(1032)/O~VI(1038),
	N~V(1238)/N~V(1242), C~IV(1548)/C~IV(1550)). 
	Ratio is listed with stronger line.  If weaker line is not
	detected above the 2$\sigma$ equivalent width limit, then that
	limit is used in the Line Ratio.}
\tablenotetext{d}{ Visual Magnitude}
\tablenotetext{e}{ $z_{ave}$ is calculated as the weighted mean of \Lya, \Lyb,
	\Lyg, O~VI(1032) \& O~VI(1038).  See text for a desciption
	of the weighting procedure.
	Standard deviation is calculated for the five lines above.
	The samples in which the system has been included are marked by 
	the letters A,B,C.}
\tablenotetext{f}{ This line had an upper limit, described in the text,
	used in place of its
	Significance Level as the
	weight for calculating $z_{ave}$. }
\tablenotetext{g}{ O~VI(1032) at \zabs = 0.9534 is blended with \Lyb at   
	\zabs=0.9669. }
\tablenotetext{h}{ Absorption due to this ion is not detected 
	at the 2$\sigma$ level at $z_{ave}$.  
	An upper limit equal to the
	2$\sigma$ equivalent width level is listed. } 
\tablenotetext{i}{ The line ratio does not fall within the 1$\sigma$ 
	error of the theoretical prediction.}
\tablenotetext{j}{ This line is blended producing excess equivalent width.}

\enddata
\end{planotable}

\begin{table*}
\begin{center}
\begin{tabular}{ccccccc}
QSO & $z_{abs}$  & 
\multispan 3 \hfil {Rest Equivalent Widths (\AA)} \hfil
& \multispan{2} \hfil Doublet Ratio \hfil \cr
& & O~VI & O~VI & C~II & Subtracted\tablenotemark{b} & Fitted\tablenotemark{c}\cr
& & 1032 & 1036-8\tablenotemark{a} & 1334 &  &\cr
\tableline
0122-0021  & 0.9534 & 0.62 & 0.44 & 0.18 & 2.00\tablenotemark{d} & 1.40\cr
           & 0.9669 & 0.36 & 0.25 & $\leq$ 0.16 & 1.44 & 1.44\cr
0454-2203 & 0.4126 & 0.60 & 0.43 & $\leq$ 0.02 & 1.40 & 1.40\cr
1206+4557 & 0.7339 & 0.16 & 0.14 & $\leq$ 0.11 & 1.14 & 1.15\cr
           & 0.9270 & 1.21 & 1.29 & $\leq$ 0.11 & 0.94 & 0.94\cr
           & 1.0828 & 1.22 & 1.04 & $\leq$ 0.10 & 1.17 & 1.18\cr
1352+0106  & 0.6675 & 0.65 & 0.85 & 0.72 & 1.98\tablenotemark{d} & 2.46\cr
1424-1150 & 0.6552 & 0.19 & 0.25 & $\leq$ 0.09 & 0.76 & 0.76\cr
           & 0.7465 & 0.31 & 0.16 & $\leq$ 0.17 & 1.94 & 1.91\cr
1634+7037 & 1.0416 & 0.21 & 0.17 & 0.10 & 1.96\tablenotemark{d}  & 1.48\cr
           & 1.1412 & 0.11 & 0.09 & $\leq$ 0.03 & 1.22 & 1.20\cr
2340-0340 & 0.6842 & 0.31 & 0.35 & 0.12 & 1.37\tablenotemark{d}  & 1.68\cr
\tablenotetext{a}{ Rest equivalent width including both O~VI(1038)
and C~II(1036).}
\tablenotetext{b}{ W(1032)/W(1038), corrected for C~II(1036) when
C~II(1334) was detected.}
\tablenotetext{c}{ W(1032)/W(1038), using W for O~VI (1038)
deblended from C~II(1036) in Table 4.}
\tablenotetext{d}{ Doublet ratio corrected for calculated C~II(1036).}

\end{tabular}
\end{center}

\tablenum{5}
\caption{Doublet Ratios} 

\end{table*}

\clearpage
 
\begin{table*}
\begin{center}
\begin{tabular}{cccccccc}

QSO & $z_{abs}$  & 
\multispan 2 \hfil {Oxygen VI\tablenotemark{b}} \hfil
& \multispan 2 \hfil {Carbon IV\tablenotemark{c}} \hfil
& \multispan 2 \hfil {Hydrogen I\tablenotemark{d}} \hfil \cr
& & Log(N) & b & Log(N) & b & Log(N) & b \cr
& & cm$^{-2}$ & \kms & cm$^{-2}$ & \kms & cm$^{-2}$ & \kms \cr
\tableline
0122-0021  & 0.9534 & 15.1$^{+0.3}_{-0.3}$ & 74$^{+70}_{-25}$ 
		    & 15.3$^{+2.3}_{-0.5}$ & 54$^{+28}_{-20}$ 
                    & 17.0$_{-0.5}^{+1.5}$ & 53$^{+10}_{-13}$ \cr
           & 0.9669 & 14.8$^{+1.3}_{-0.2}$ & 45$_{-26}^{+100}$ 
		    & 13.9\tablenotemark{e} & \nodata
		    & 16.1$_{-0.2}^{+0.4}$ & 49$^{+8}_{-9}$ \cr
0454-2203 & 0.4126  & 15.1$^{+0.2}_{-0.2}$ & 70$_{-20}^{+39}$ 
		    & \lq13.5\tablenotemark{g} & \nodata
		    & \nodata\tablenotemark{f} & \nodata\tablenotemark{f} \cr
1206+4557 & 0.7339  & \nodata\tablenotemark{f} & \nodata\tablenotemark{f} 
		    & 13.7\tablenotemark{e} & \nodata
		    & \nodata\tablenotemark{f} & \nodata\tablenotemark{f} \cr
           & 0.9270 & \nodata\tablenotemark{f} & \nodata\tablenotemark{f} 
		    & 15.0$_{-0.1}^{+0.1}$ & 223$^{+45}_{-1}$  
		    & 18.7$^{+0.9}_{-0.7}$ & 86$^{+9}_{-10}$ \cr
           & 1.0828 & 15.7$^{+0.2}_{-0.1}$ & 106$^{+10}_{-10}$ 
		    & \nodata\tablenotemark{f} & \nodata\tablenotemark{f} 
		    & \nodata\tablenotemark{f} & \nodata\tablenotemark{f} \cr
1352+0106  & 0.6675 & 14.7\tablenotemark{e} & \nodata
		    & 15.3$_{-0.2}^{+0.7}$ & 74$^{+13}_{-18}$
		    & \nodata\tablenotemark{f} & \nodata\tablenotemark{f} \cr 
1424-1150 & 0.6552  & \nodata\tablenotemark{f} & \nodata\tablenotemark{f} 
		    & \nodata\tablenotemark{f} & \nodata\tablenotemark{f} 
		    & 17.8$^{+2.0}_{-0.9}$ & 49$^{+6}_{-16}$ \cr
           & 0.7465 & 14.4\tablenotemark{e} & \nodata
		    & \lq13.5\tablenotemark{g} & \nodata
		    & 13.7\tablenotemark{e} & \nodata \cr
1634+7037 & 1.0416  & 14.8$_{-0.5}^{+1.3}$ & 20$^{+50}_{-10}$ 
		    & 14.5$_{-0.1}^{+0.1}$ & 40$^{+11}_{-7}$ 
		    & 16.6$^{+0.2}_{-0.1}$ & 63$^{+3}_{-3}$ \cr
           & 1.1412 & 14.6$_{-0.5}^{+1.7}$ & 10$^{+30}_{-4}$ 
		    & \nodata & \nodata
		    & 15.1$_{-0.2}^{+0.2}$ & 31$^{+3}_{-3}$ \cr
2340-0340 & 0.6842  & 14.6$^{+1.4}_{-0.2}$ & 45$_{-32}^{+\infty}$ 
		    & \nodata & \nodata
		    & \nodata\tablenotemark{f} & \nodata\tablenotemark{f} \cr
\tablenotetext{a}{ Values are calculated with two equivalent width
measurements and their 1$\sigma$ errors.}
\tablenotetext{b}{ From lines $\lambda\lambda1032,1038$.}
\tablenotetext{c}{ From lines $\lambda\lambda1548,1550$.}
\tablenotetext{d}{ From lines $\lambda1215$, $\lambda1025$.}
\tablenotetext{e}{ Doublet ratio suggests linear absorption. 
Column density estimate is 
based on the stronger member. } 
\tablenotetext{f}{ Absorption lines are too saturated to be
effectively analyzed in with the doublet ratio method.}
\tablenotetext{g}{ Neither line was detected above 2$\sigma$.  Upper
limit is placed on column density assuming linear absorption. }

\end{tabular}
\end{center}

\tablenum{6}
\caption{Column Densities and Dispersion Velocities\tablenotemark{a}} 

\end{table*}

\clearpage
 
\begin{table*}
\begin{center}
\begin{tabular}{ccccccccc}
Object & \zabs & W(1032) & W(1025) & $R_{OB}$ & $\sigma_{R_{OB}}$ & 
		W(1548) & $R_{OC}$ & $\sigma_{R_{OC}}$ \cr
\tableline
0122-0021 &  0.9534 & 1.21 & 1.37 & 0.88 & 0.10 & 1.99 & 0.56 & 0.07 \cr
             & 0.9669 & 0.71 & 0.80 & 0.89 & 0.16  & 0.63 & 1.13 & 0.36 \cr
0454-2203 & 0.4126 & 0.85 & 0.68 & 1.25 & 0.14  & $\leq0.07$\tablenotemark{a} & 
		$\geq12.1$ & ... \cr
1206+4557 & 0.7339 & 0.28 & $\leq0.85$\tablenotemark{b}
 & 0.33 & 0.09 & 0.37 & 0.76 & 0.25 \cr
             & 0.9270 & 2.32 & 3.00 & 0.77 & 0.03 & 4.44 & 0.53 & 0.02 \cr
             & 1.0828 & 2.54 & 1.26 & 2.02 & 0.08 & 0.61 & 4.17 & 0.71 \cr
1352+0106  & 0.6675 & 1.09 & 2.18 & 0.50 & 0.05 & 2.22 & 0.49 & 0.04 \cr
1424-1150 & 0.6552 & 0.31 & 1.24 & 0.25 & 0.08 & 0.65 & 0.48 & 0.19 \cr
	  & 0.7465 & 0.54 & $\leq0.17$\tablenotemark{c} & $\geq3.18$ 
	  	& ... & $\leq0.27$\tablenotemark{a} & $\geq2.00$ & ... \cr
1634+7037  & 1.0416 & 0.42 & 1.41 & 0.33 & 0.04 & 1.04 & 0.40 & 0.05 \cr
           & 1.1412 & 0.24 & 0.17 & 1.41 & 0.48 & $\leq0.07$\tablenotemark{a} & 
		$\geq3.4$ & ... \cr
2340-0340 & 0.6842 & 0.53 & 1.88 & 0.28 & 0.08 & $\leq0.32$\tablenotemark{a} & 
		$\geq1.7$ & ... \cr
\tablenotetext{a} { C~IV(1548) is not detected, an upper limit 2$\sigma$
equivalent width is listed. }
\tablenotetext{b}{ \Lyb is apparently blended with another \Lya line;
an upper limit of the
equivalent width is given by its corresponding \Lya.}
\tablenotetext{c} { \Lyb is not detected, an upper limit 2$\sigma$
equivalent width is listed. }

\end{tabular}
\end{center}

\tablenum{7}
\caption{Ion Ratios} 

\end{table*}

\clearpage

\begin{figure}
\figurenum{1}
\caption{ Number of chance coincidences of new redshift systems with 
two lines (O~VI pairs),
three lines (\Lya and O~VI), and four lines (\Lya, \Lyb and O~VI).
We used a random distribution of lines, and accpeted lines into a system
if the redshift for each line fell within $\Delta z = 0.0005$ of all the
others.  
We see that the number of false three lined systems 
is about 10\% that of the two lined systems.
Each HST spectrum covers 700 -- 1000~\AA\ , the real lines density is
80.0 per 1000 \ang in the Lyman-$\alpha$ forest, and real O~VI
systems occur about 1 per 1000~\AA\ in the rest frame.  Therefore,
we require at least 3 lines per system in order to avoid excessive
coincidences.}
\end{figure}

\begin{figure}
\figurenum{2}
\caption{ The spectra of 11 QSOs plotted with the fitted continuum and 1$\sigma$
error.  Lines identified with Oxygen VI absorption systems are labeled (see
Table 4).  
Labels of the same height indicate lines of a common system.  }
\end{figure}

\begin{figure}
\figurenum{3}
\caption{ Plots
of Lyman-$\alpha$, Lyman-$\beta$, O~VI, and C~IV absorption in velocity
space for each of the 12 systems in which O~VI was identified.  Zero velocity
corresponds to the \zabs indicated for each system.  Flux scales are
different depending on the strength of absorption.}
\end{figure}

\begin{figure}
\figurenum{4}
\caption{ Calculated
O~VI Doublet Ratio versus Column Density for a single absorbing
cloud.  Our search was sensitive to the right
of the vertical dashed line at log(N$_{OVI}$) $>$ 14.5,
which corresponds to W(1038) $>$ 0.21 \AA.}
\end{figure}

\begin{figure}
\figurenum{5}
\caption{ Equivalent wdith ratios for 
the 12 systems in Sample A.  Ratios of equivalent widths
and the 1$\sigma$ error are calculated in each system where absorption
was detected at the respective wavelengths.  Lower limits are shown
with arrows in the systems where C~IV or \Lyb are not detected.  }
\end{figure}

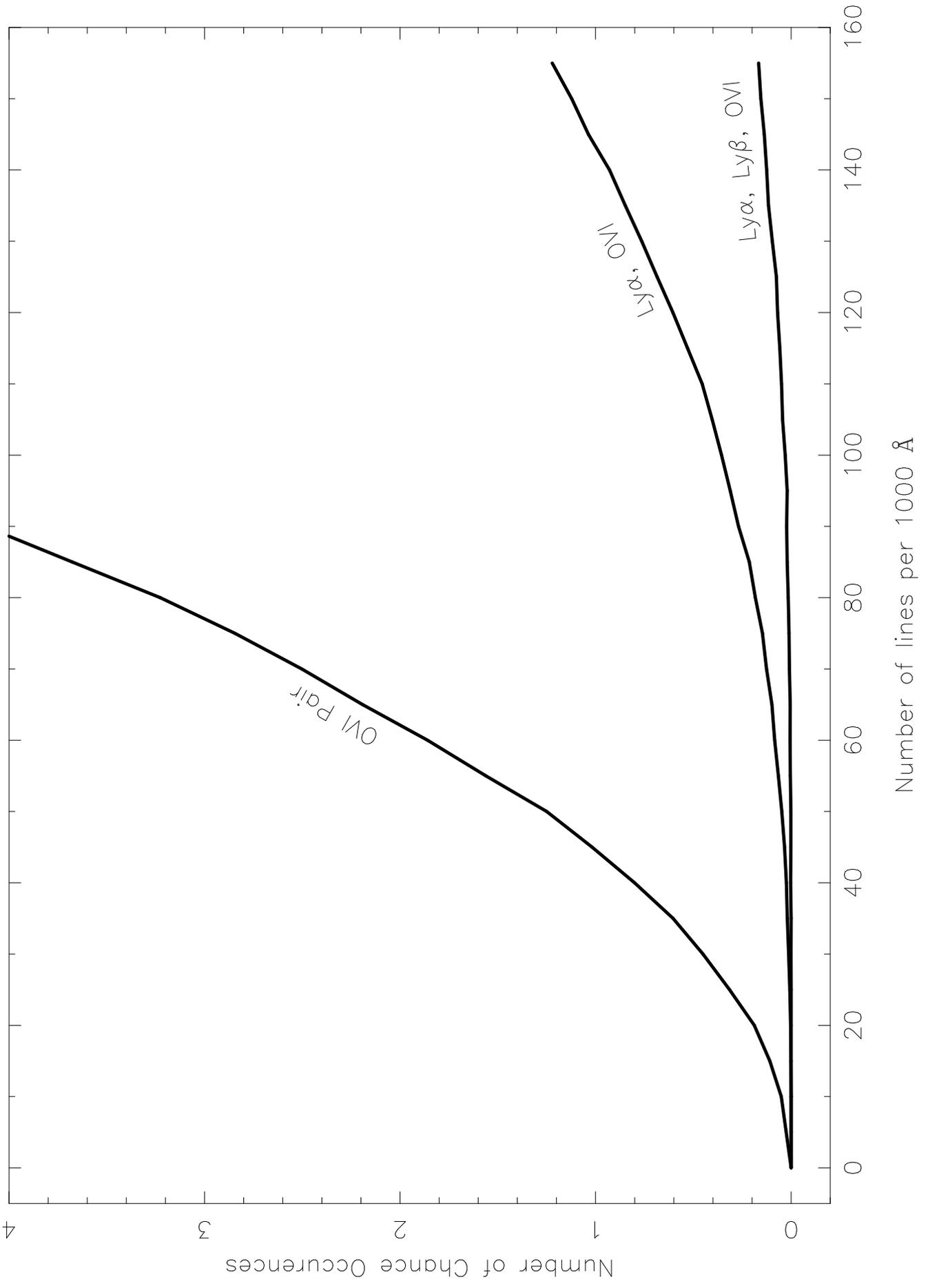
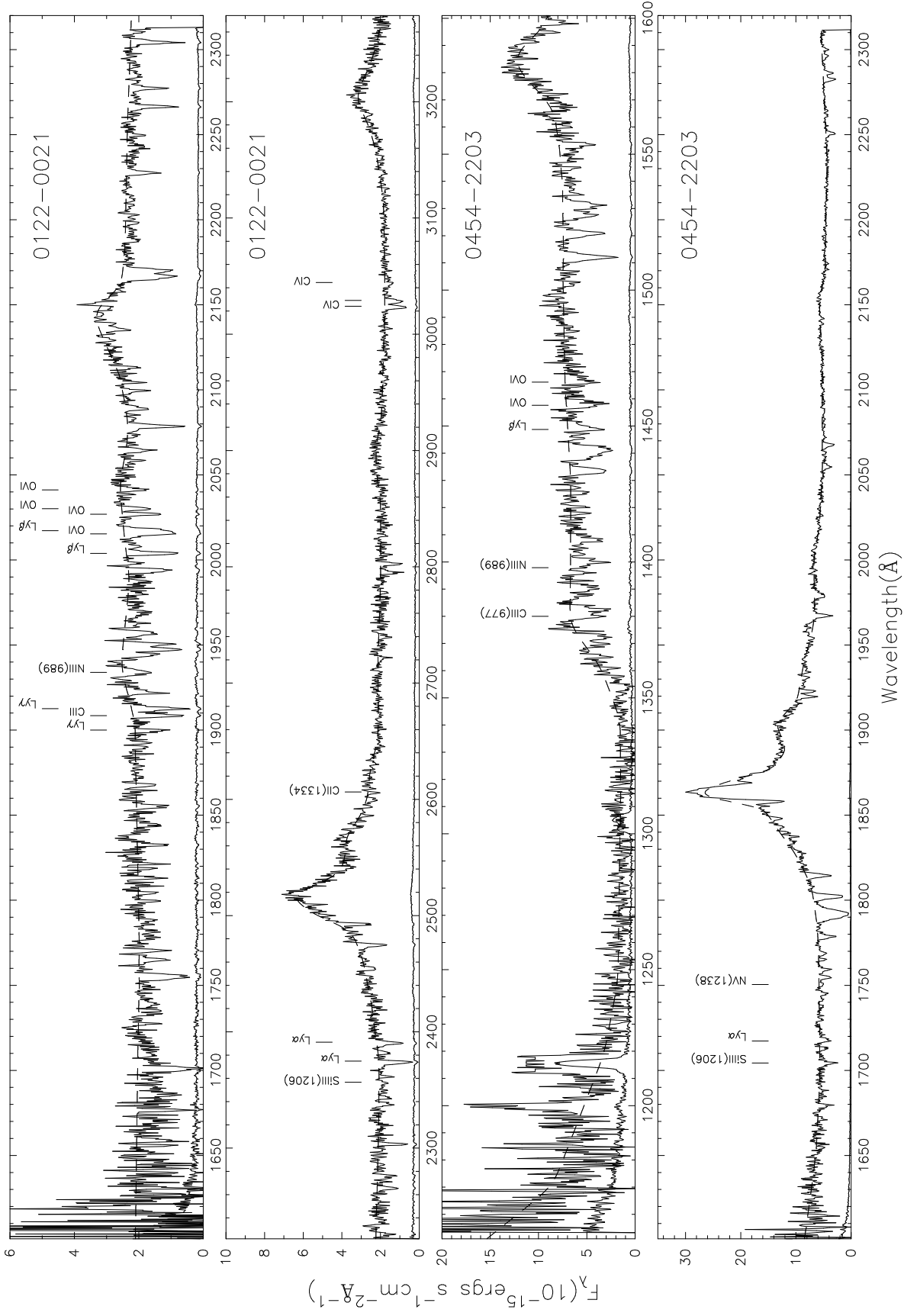
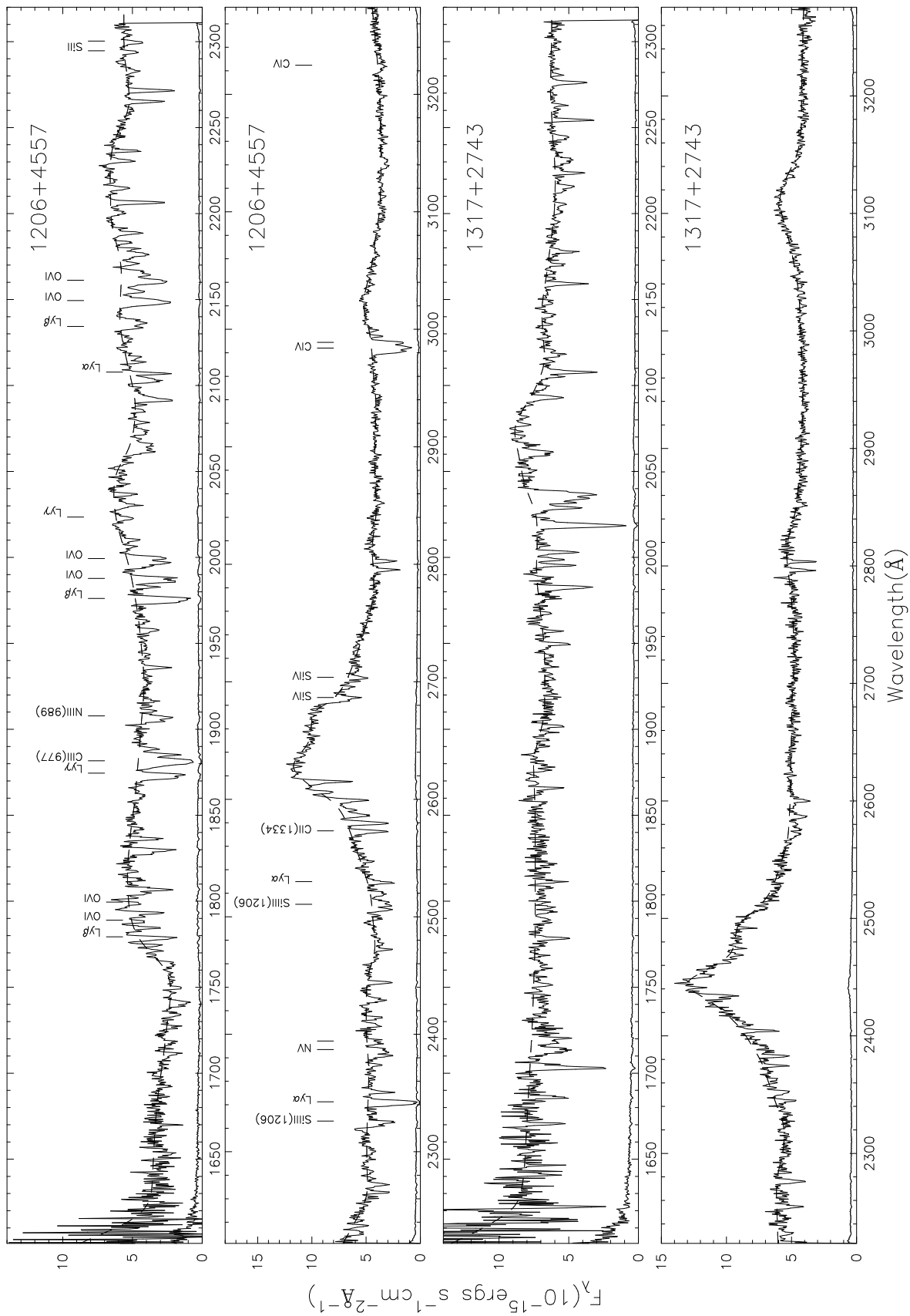
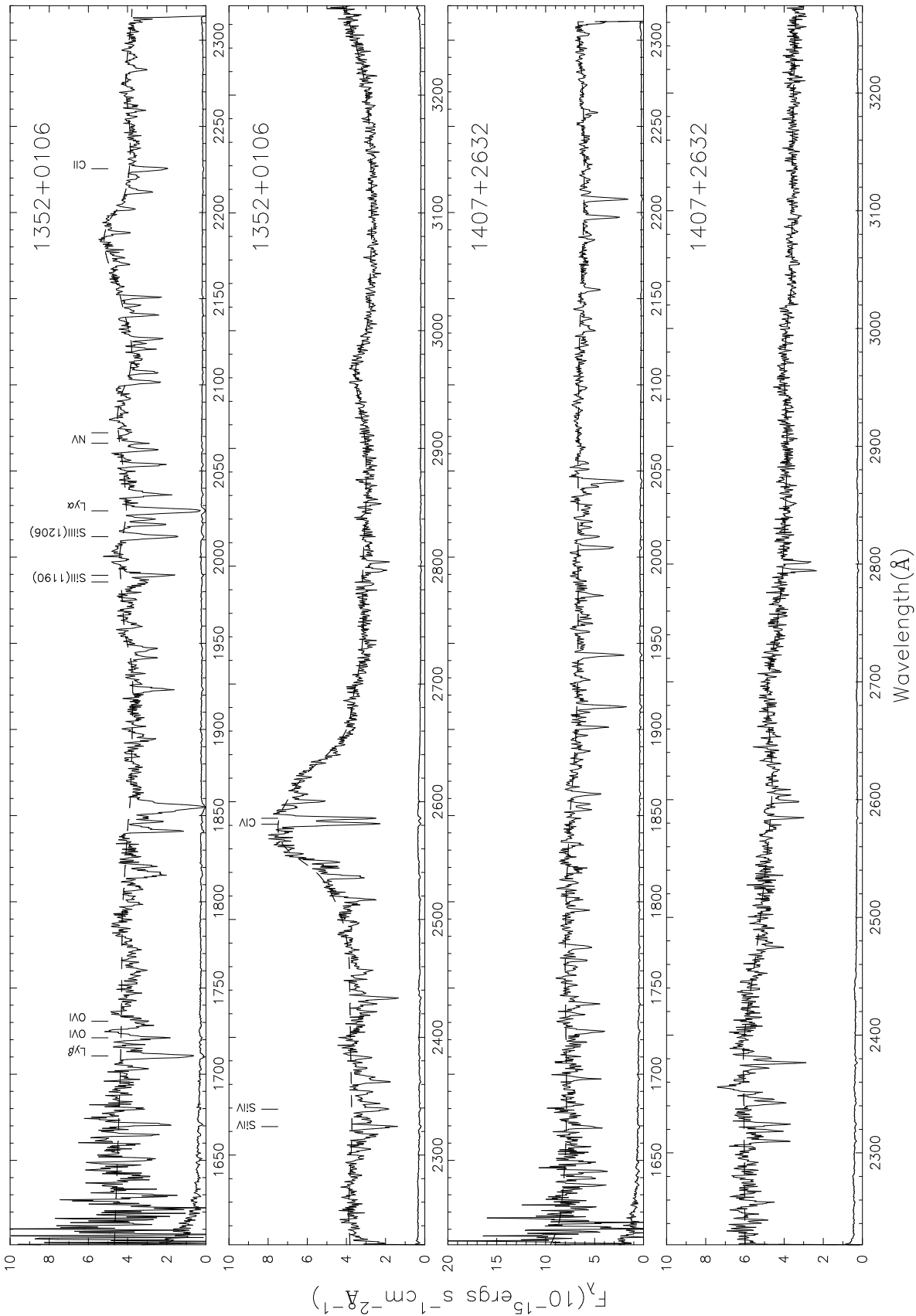
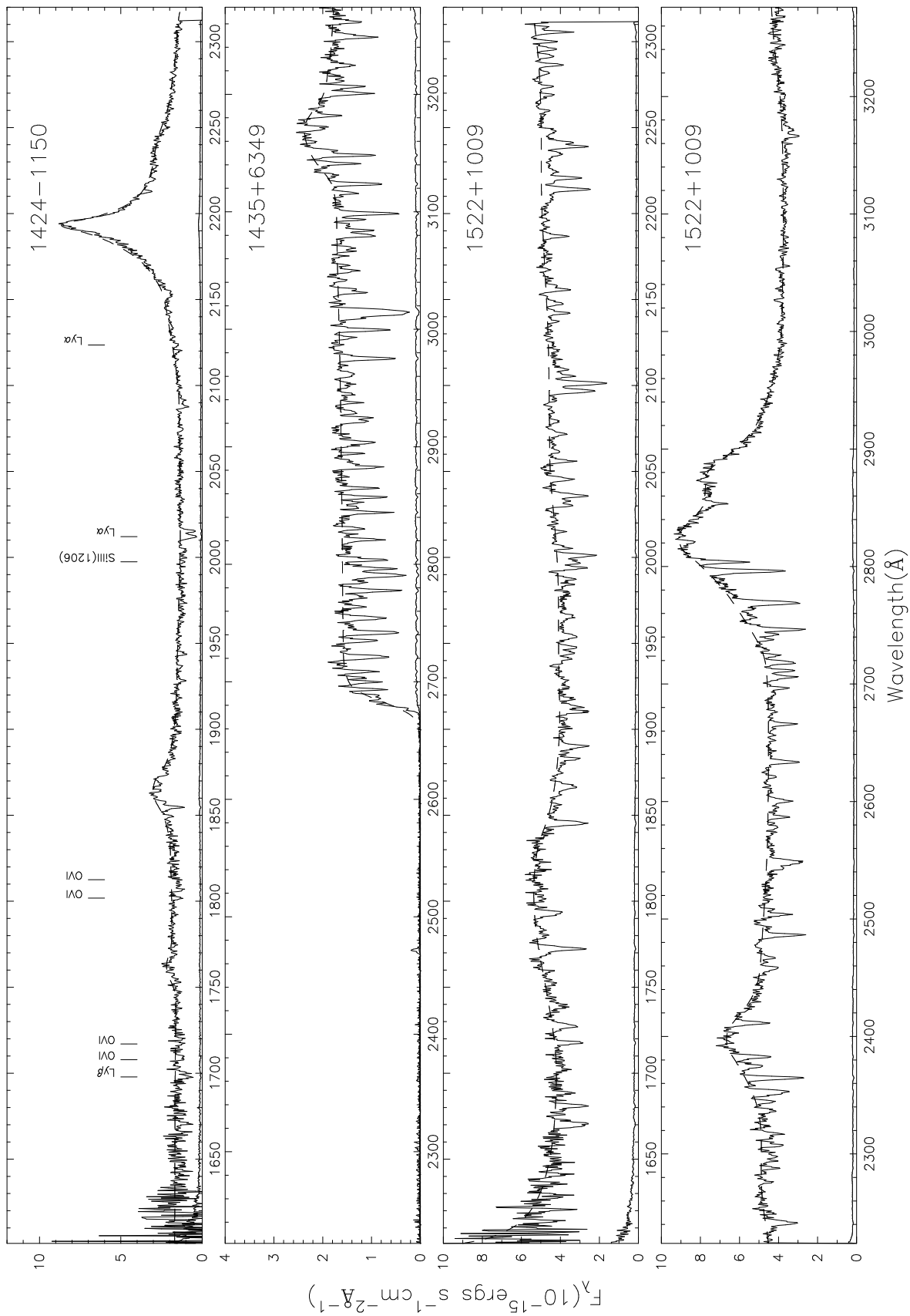
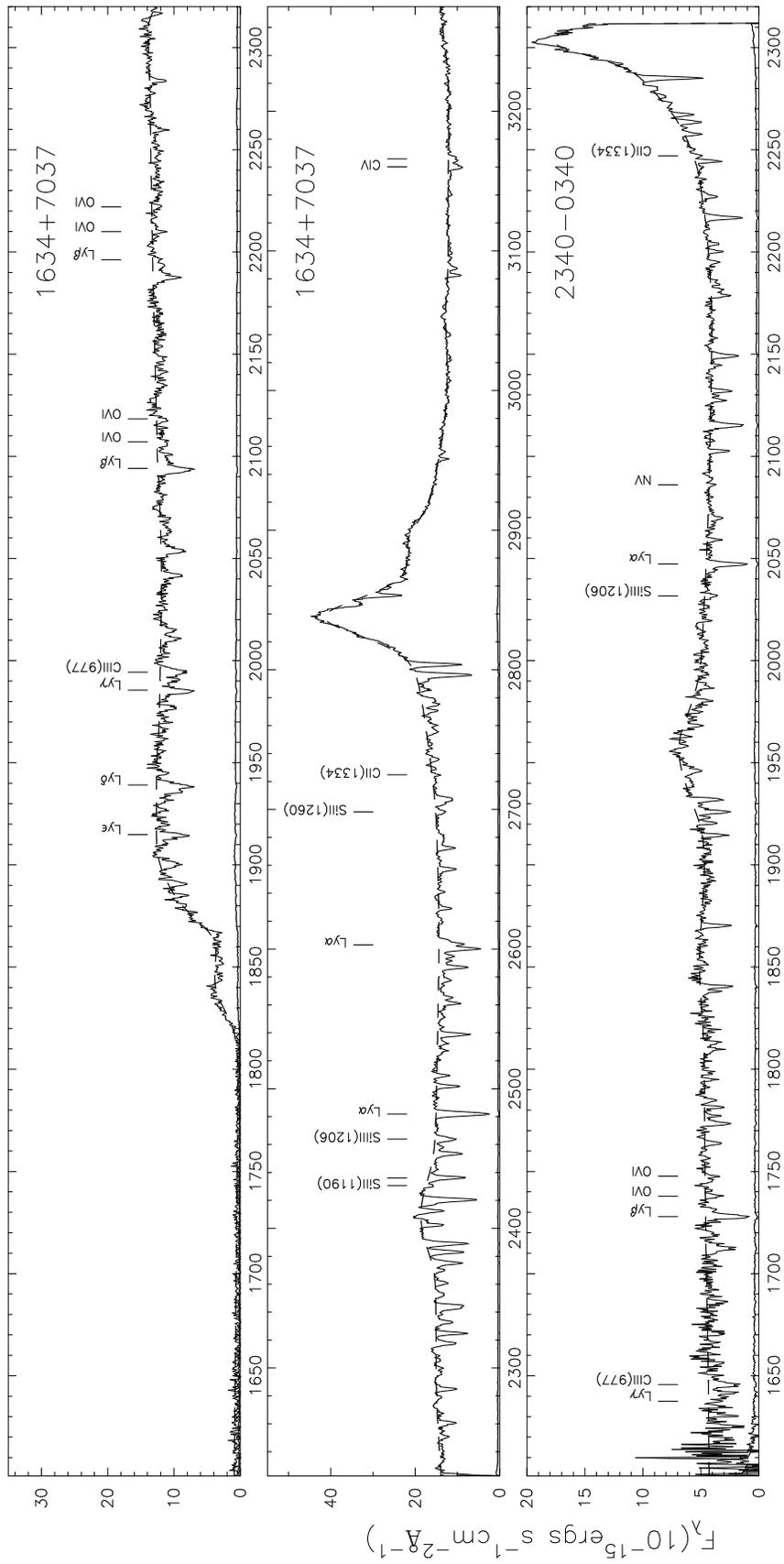
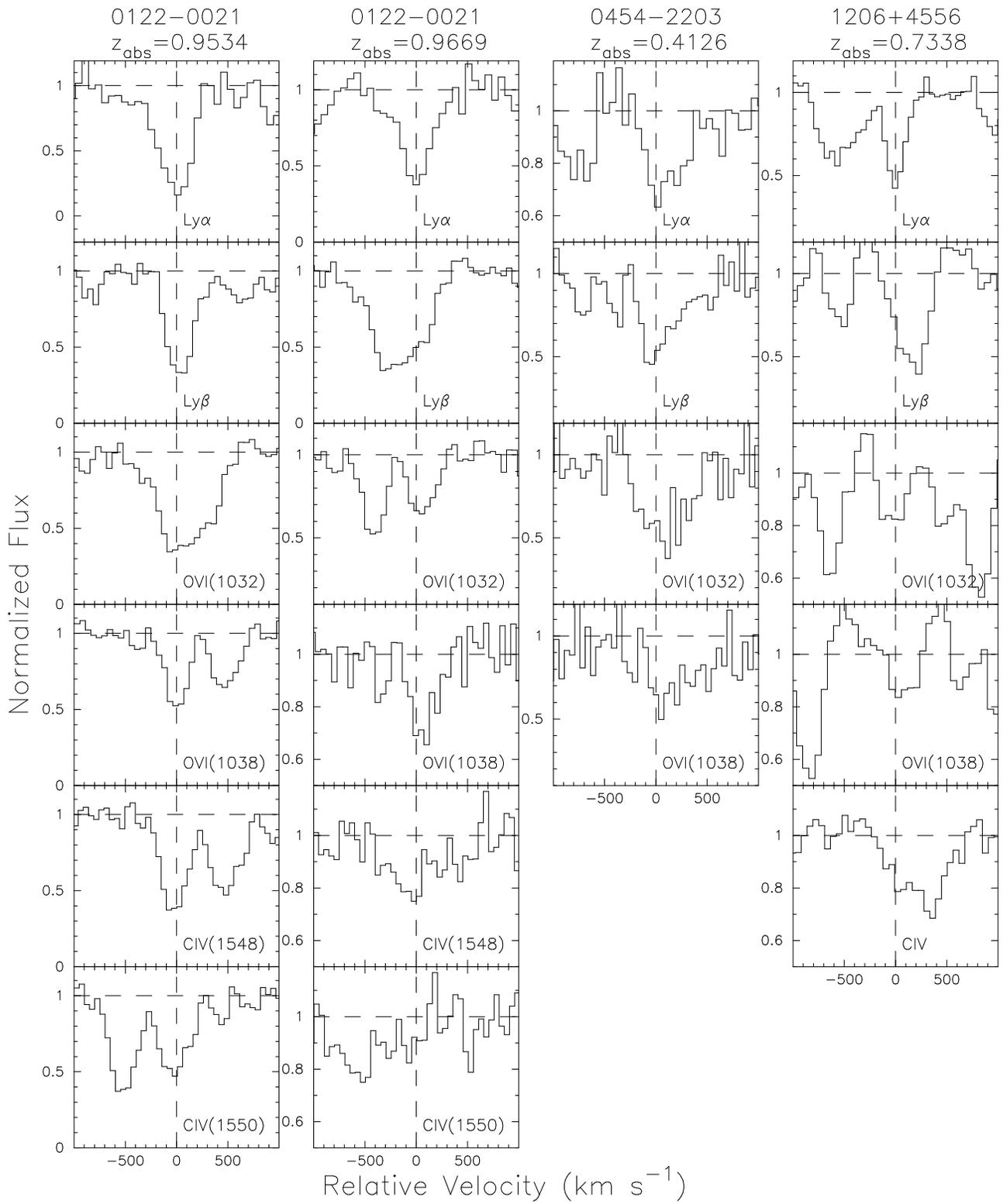
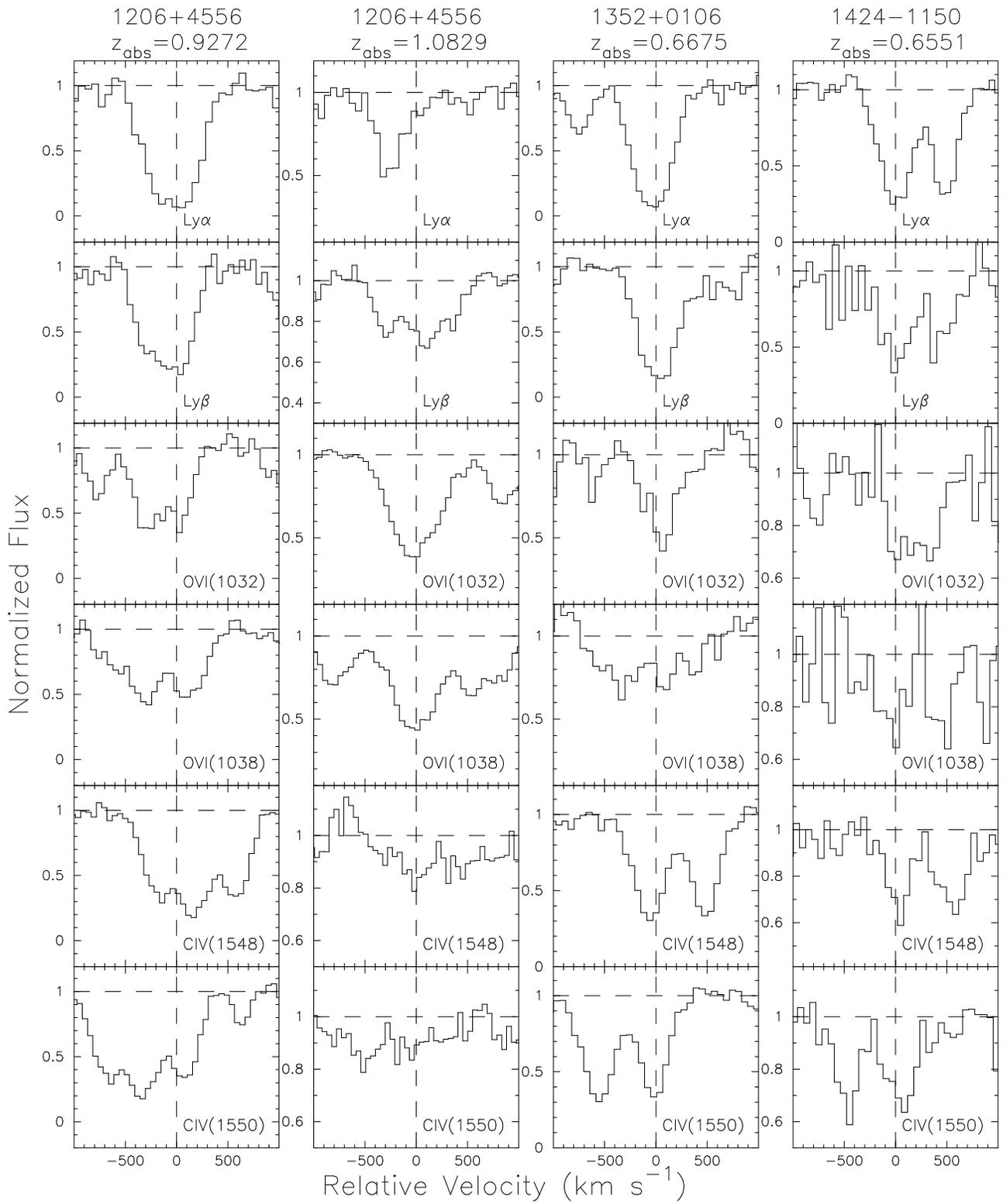
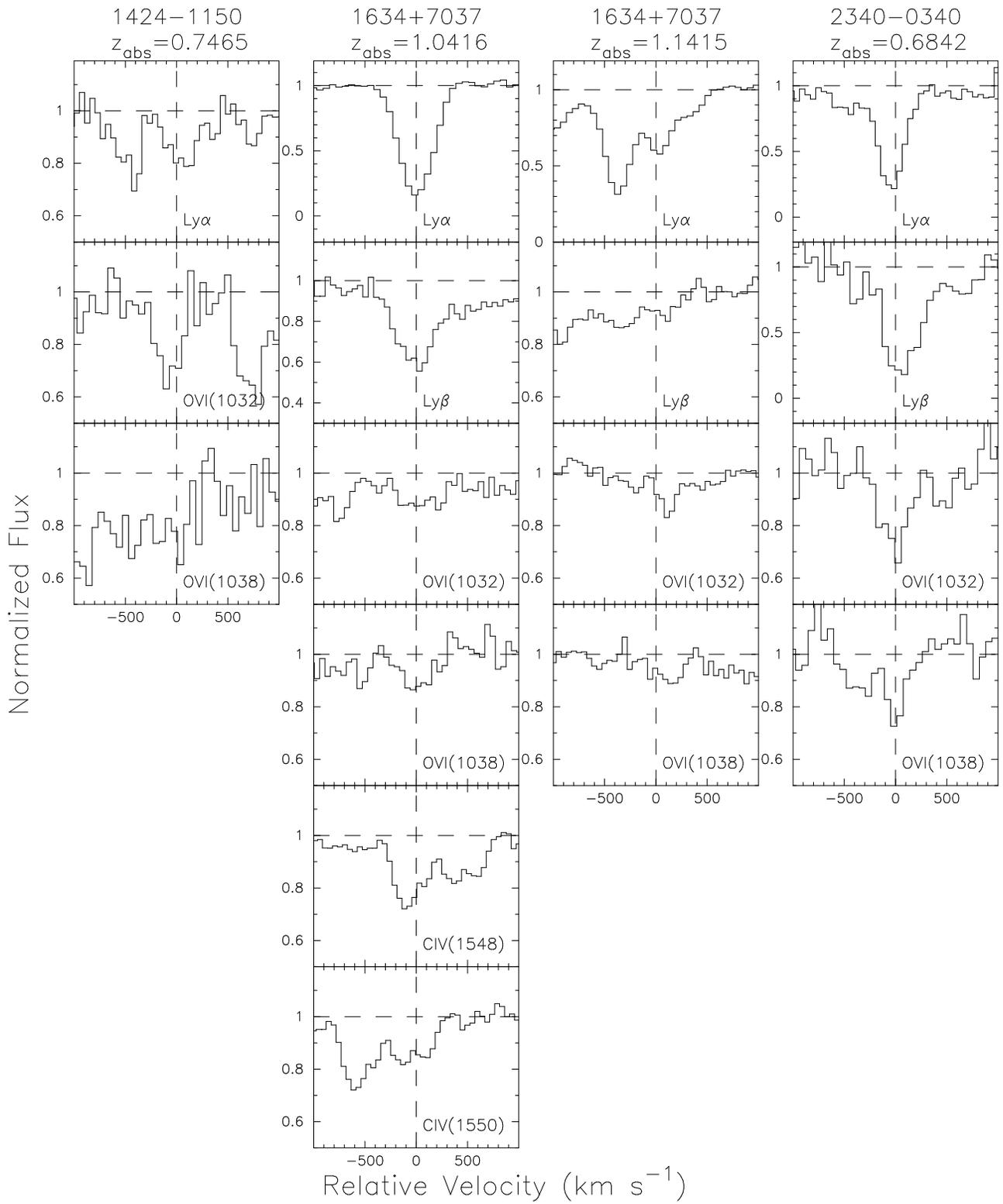
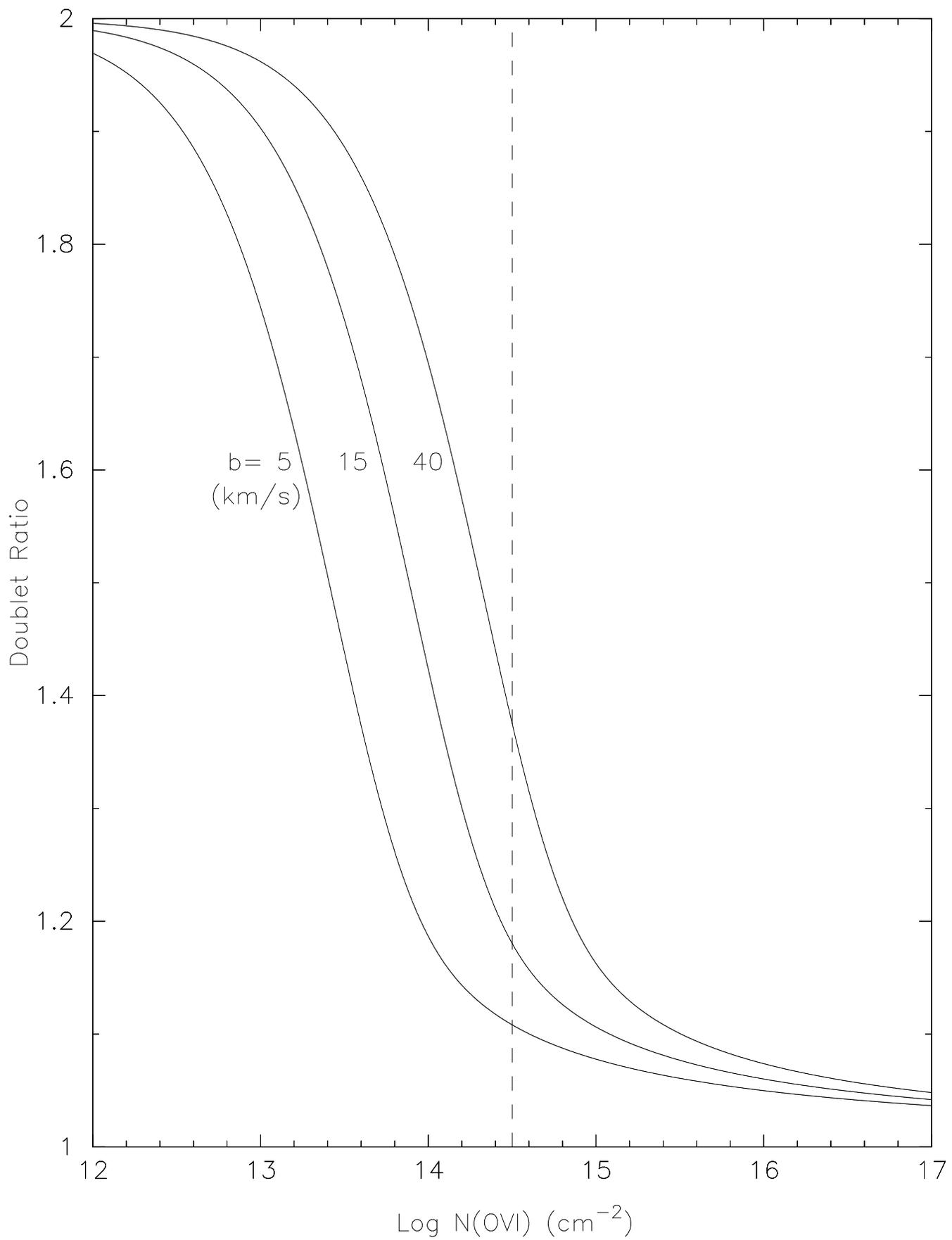
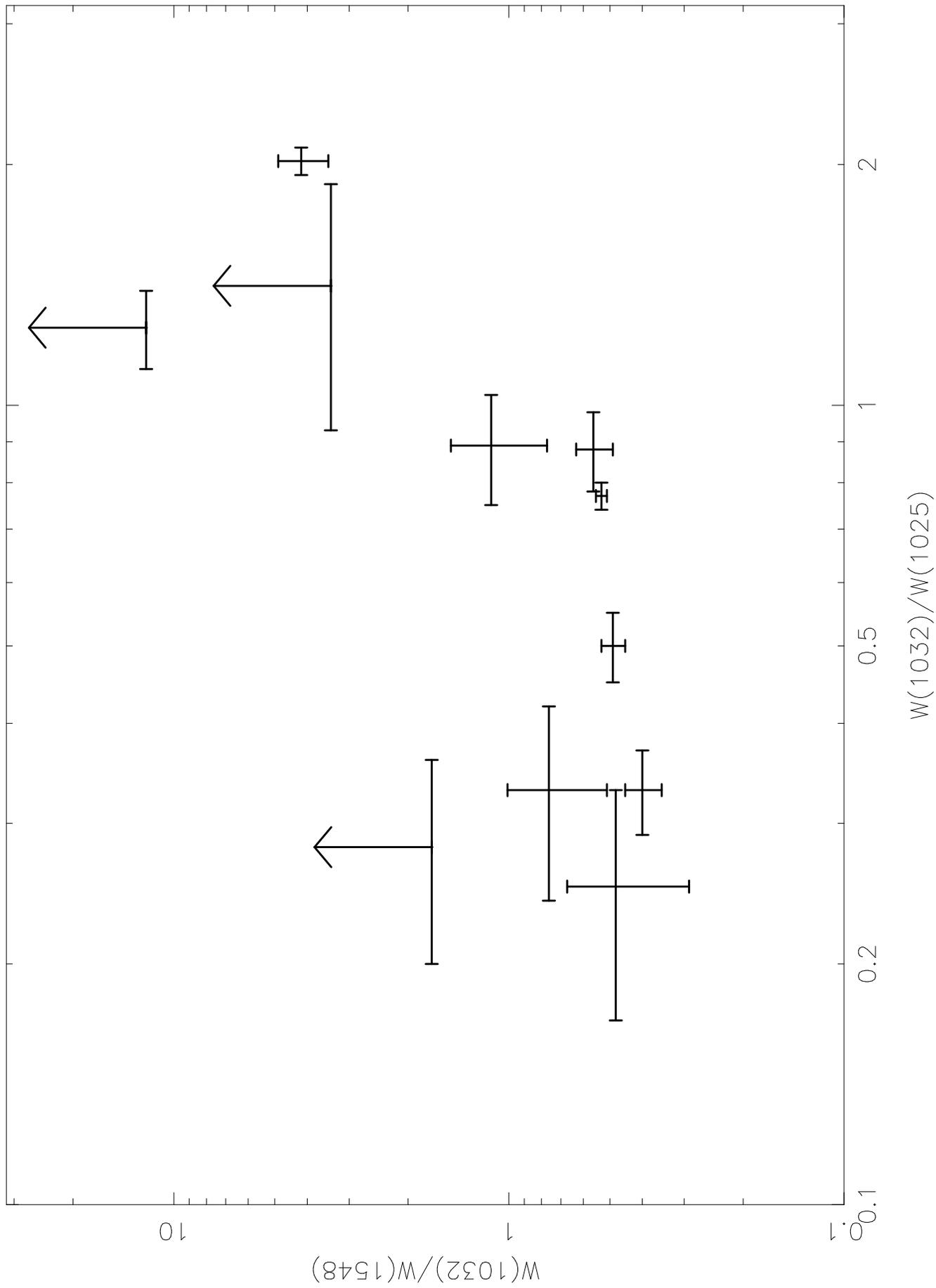
\end{document}